\documentclass[useAMS,usenatbib]{mn2e}

\usepackage{times}

\usepackage{epsfig}
\usepackage{graphicx}              
\usepackage{afterpage}
\usepackage{deluxetable}
\usepackage{hyperref}
\usepackage{times}

\title[Fundamental Parameters of Exoplanet Host Stars]
{Stellar Diameters and Temperatures V. \\Eleven Newly Characterized Exoplanet Host Stars}
\author[von Braun, Boyajian, et al.]
{\parbox{\textwidth}{Kaspar von Braun$^{1,2}$\thanks{E-mail: braun@mpia.de},
Tabetha S. Boyajian$^{3}$,
Gerard T. van Belle$^{4}$,
Stephen R. Kane$^{9}$,
Jeremy Jones$^{5}$,
Chris Farrington$^{7}$,
Gail Schaefer$^{7}$,
Norm Vargas$^{7}$,
Nic Scott$^{7}$,
Theo A. ten Brummelaar$^{7}$,
Miranda Kephart$^{3}$,
Douglas R. Gies$^{5}$,
David R. Ciardi$^{6}$, 
Mercedes L\'{o}pez-Morales$^{8}$,
Cassidy Mazingue$^{2}$,
Harold A. McAlister$^{7}$,
Stephen Ridgway$^{10}$,
P. J. Goldfinger$^{7}$,
Nils H. Turner$^{7}$,
and Laszlo Sturmann$^{7}$\\
}
\vspace{0.3cm}\\
$^{1}$Max-Planck-Institute for Astronomy (MPIA), K\"{o}nigstuhl 17, 69117 Heidelberg, Germany\\
$^{2}$Mirasol Institute, Munich, Germany\\
$^{3}$Yale University, New Haven, CT, USA\\
$^{4}$Lowell Observatory, Flagstaff, USA\\
$^{5}$Center for High Angular Resolution Astronomy and Department of Physics and Astronomy, \\Georgia State University, P. O. Box 5060, Atlanta, GA 30302-4106, USA\\
$^{6}$NASA Exoplanet Science Institute, California Institute of Technology, MC 100-22, Pasadena, CA 91125, USA\\
$^{7}$The CHARA Array, Mount Wilson Observatory, Mount Wilson, CA 91023, USA\\
$^{8}$CfA, Cambridge, MA, USA\\
$^{9}$Department of Physics and Astronomy, San Francisco State University, 1600 Holloway Ave., San Francisco, CA 94132, USA\\
$^{10}$NOAO, Tucson, AZ, USA
}

\begin{document}

\date{Accepted 2013 December 3. Received 2013 November 21; in original form September 13.}


\maketitle

\label{firstpage}


\begin{abstract}

\noindent
We use near-infrared interferometric data coupled with trigonometric parallax values and spectral energy distribution fitting to directly determine stellar radii, effective temperatures, and luminosities for the exoplanet host stars 61~Vir, $\rho$~CrB, GJ~176, GJ~614, GJ~649, GJ~876, HD~1461, HD~7924, HD~33564, HD~107383, and HD~210702. Three of these targets are M dwarfs. Statistical uncertainties in the stellar radii and effective temperatures range from 0.5\% -- 5\% and from 0.2\% -- 2\%, respectively. For eight of these targets, this work presents the first directly determined values of radius and temperature; for the other three, we provide updates to their properties. The stellar fundamental parameters are used to estimate stellar mass and calculate the location and extent of each system's circumstellar habitable zone. Two of these systems have planets that spend at least parts of their respective orbits in the system habitable zone: two of GJ~876's four planets and the planet that orbits HD~33564. We find that our value for GJ~876's stellar radius is more than 20\% larger than previous estimates and frequently used values in the astronomical literature.

\end{abstract}


\begin{keywords}
infrared: stars -- planetary systems -- stars: fundamental parameters (radii, temperatures, luminosities) -- stars: individual (61~Vir, $\rho$~CrB, GJ~176, GJ~614, GJ~649, GJ~876, HD~1461, HD~7924, HD~33564, HD~107383, HD~210702) -- stars: late-type -- techniques: interferometric
\end{keywords}


\section{Introduction}\label{sec:introduction}

In the characterization of exoplanetary systems, knowledge of particularly the stellar radius and temperature are of paramount importance as they define the radiation environment in which the planets reside, and they enable the calculation of the circumstellar habitable zone's (HZ) location and boundaries. Furthermore, the radii and densities of any transiting exoplanets, which provide the deepest insights into planet properties such as exoatmospheric studies or the studies of planetary interior structures, are direct functions of the radius and mass of the respective parent star. Recent advances in sensitivity and angular resolution in long-baseline interferometry at wavelengths in the near-infrared and optical range have made it possible to circumvent assumptions of stellar radius by enabling direct measurements of stellar radius and other astrophysical properties for nearby, bright stars \citep[e.g., ][and references therein]{bai08,bai08a,bai09,bai10,van09,von11c,von11a,von12,boy12a,boy12b,boy13,hub12}. 


In this paper, we present interferometric observations (\S \ref{sec:observations}) that, in combination with trigonometric parallax values, produce directly determined stellar radii for eleven exoplanet host stars\footnote{This includes HD~107383 whose companion's minimum mass is around 20 Jupiter masses (Table \ref{tab:parameters}) and could thus be considered a brown dwarf.}, along with estimates of their stellar effective temperatures based on literature photometry (\S \ref{sec:properties}). We use these empirical stellar parameters to calculate stellar masses/ages where possible, and the locations and boundaries of the system HZs (\S \ref{sec:calculated}). We discuss the implications for all the individual systems in \S \ref{sec:discussion} and conclude in \S \ref{sec:conclusion}.


\section{Data} \label{sec:data}

In order to be as empirical as possible in the calculation of the stellar parameters of our targets, we rely on our interferometric observations to obtain angular diameters (\S \ref{sec:observations}), and we fit empirical spectral templates to literature photometry to obtain bolometric flux values (\S \ref{sec:fbol}).


\subsection{Interferometric Observations}\label{sec:observations}

Our observational methods and strategy are described in detail in \S 2.1 of \citet{boy13}. We repeat aspects specific to the observations of the individual targets below. 

The Georgia State University Center for High Angular Resolution Astronomy (CHARA) Array \citep{ten05} was used to collect our interferometric observations of exoplanet hosts in $J, H$, and $K'$ bands with the CHARA Classic beam combiner in single-baseline mode. The data were taken between 2010 and 2013 in parallel with our interferometric survey of main-sequence stars \citep{boy12b,boy13}. Our requirement that any target be observed on at least two nights with at least two different baselines serves to eliminate or reduce systematic effects in the observational results \citep{von12,boy13}.  We note that were not able to adhere this strategy for HD~107383, which was only observed during one night due to weather constraints during the observing run.

\begin{figure*}
  \begin{center}
    \begin{tabular}{cc}
      \includegraphics[angle=0,width=8.2cm]{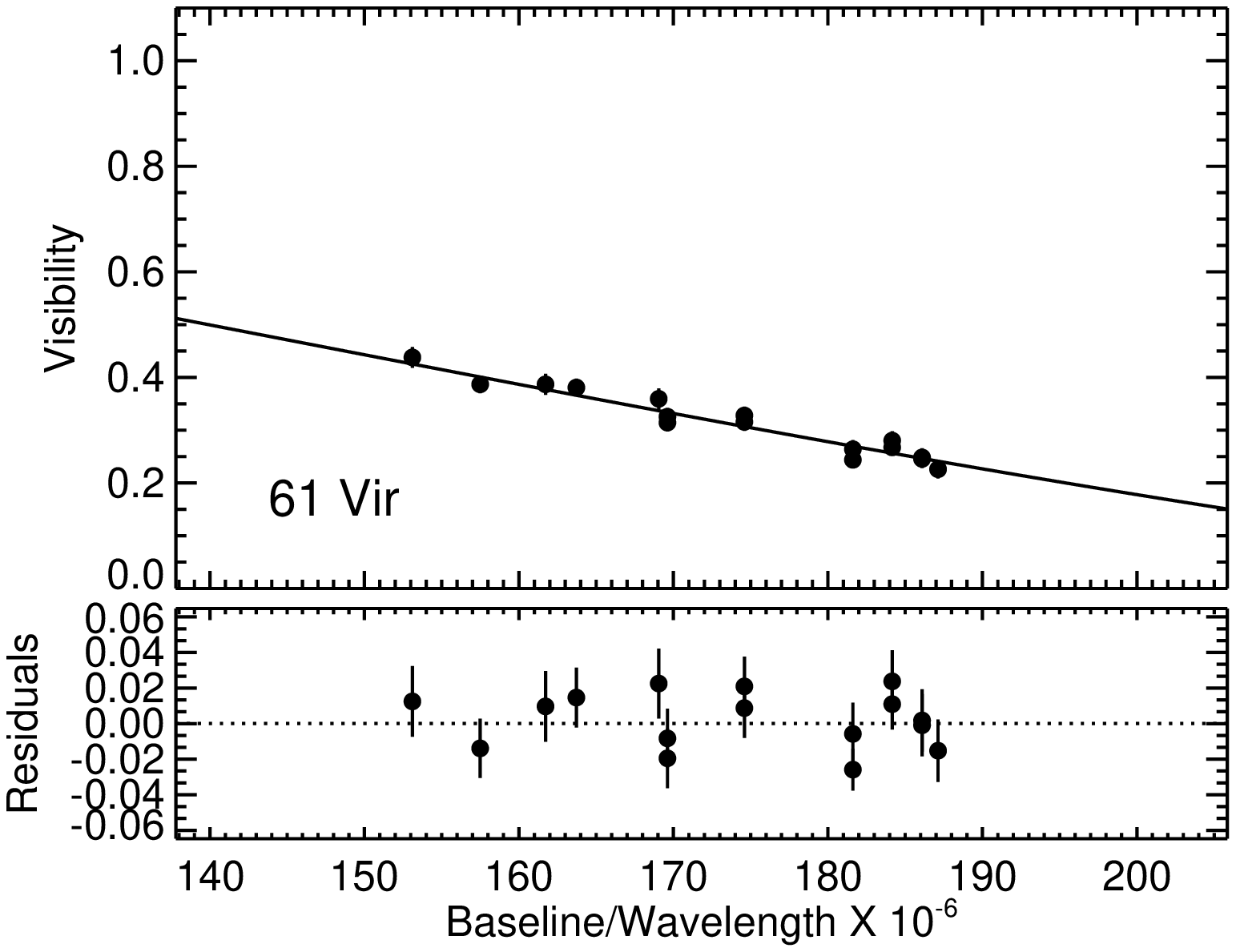} &
      \includegraphics[angle=0,width=8.2cm]{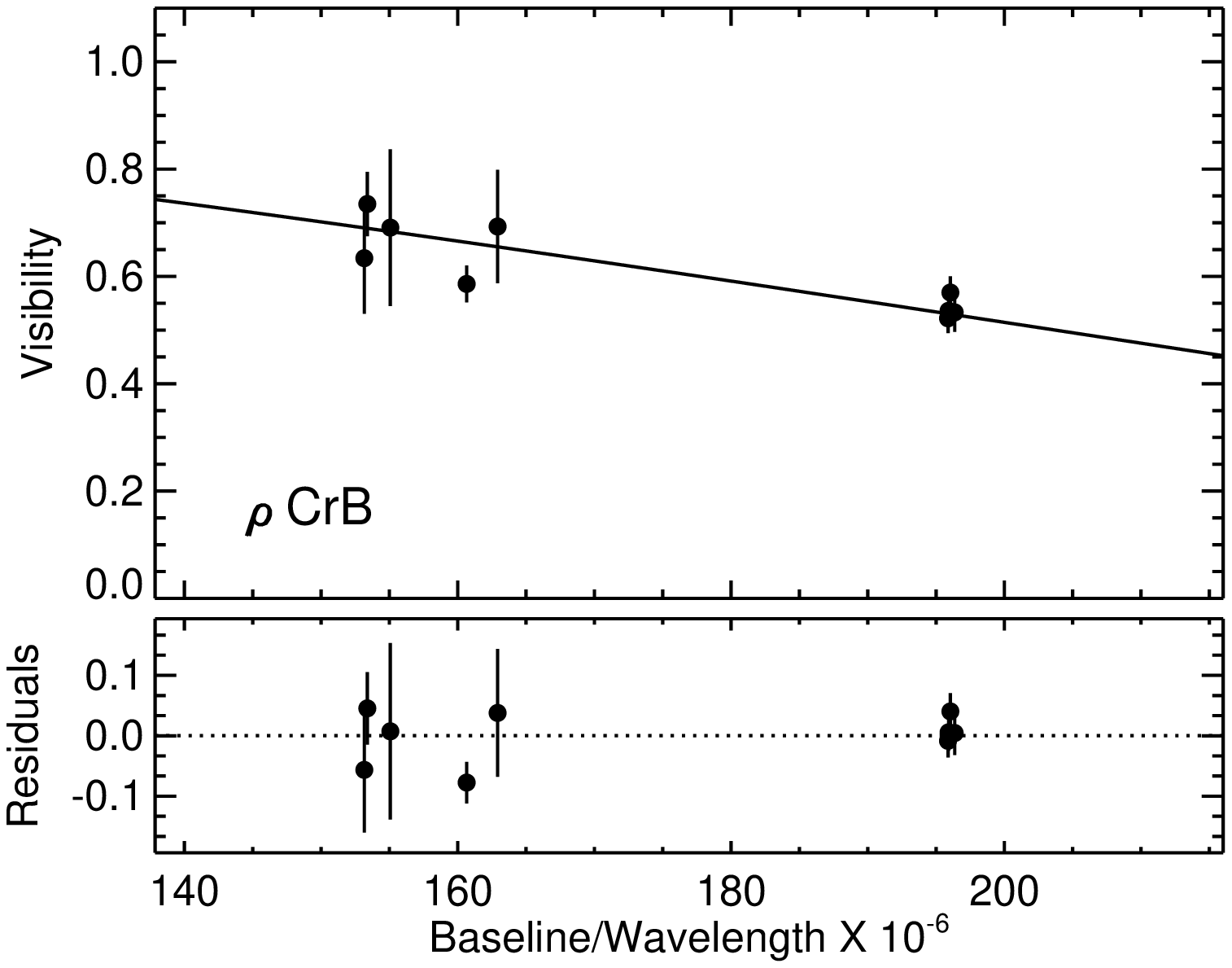} \\
      \includegraphics[angle=0,width=8.2cm]{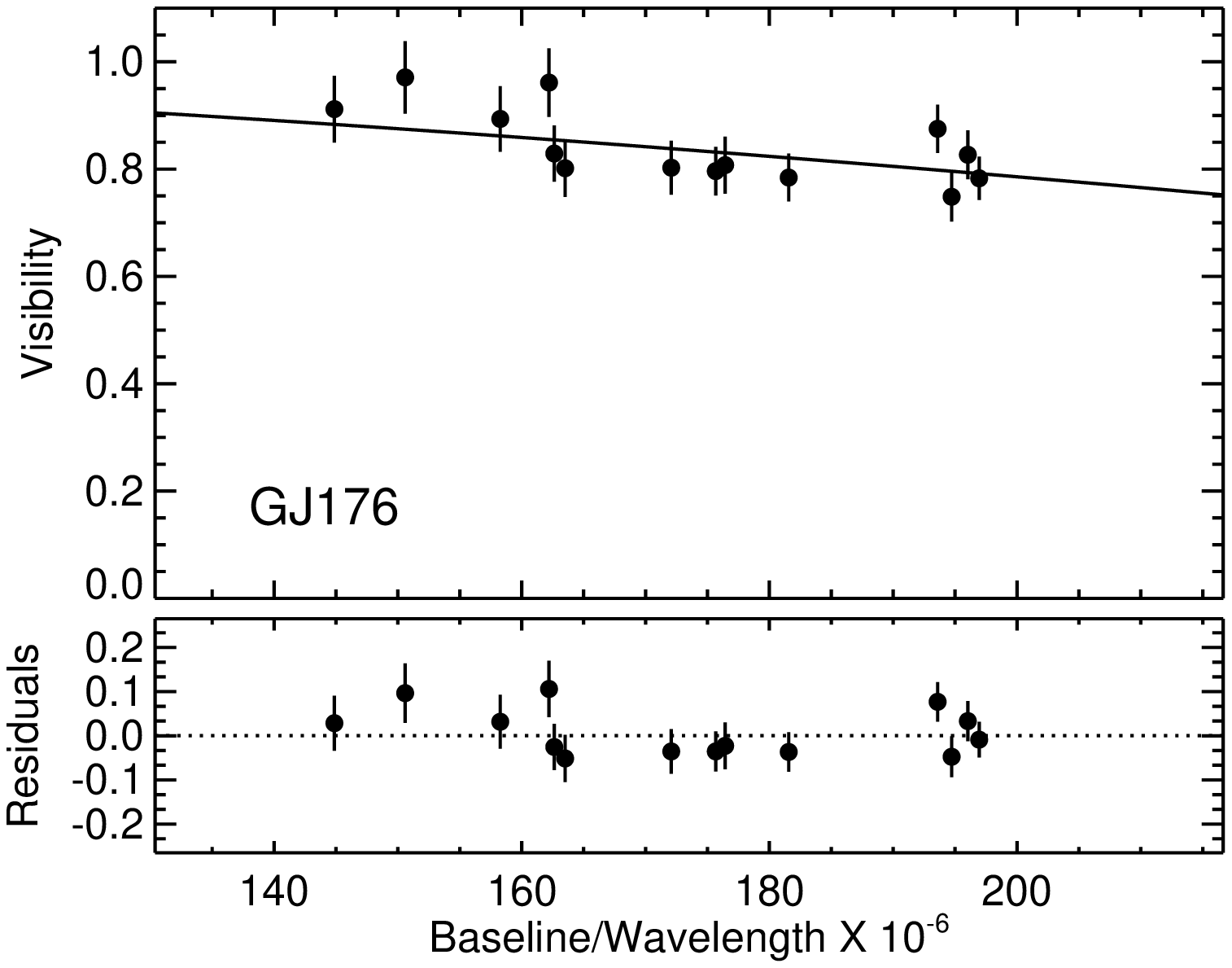} &
      \includegraphics[angle=0,width=8.2cm]{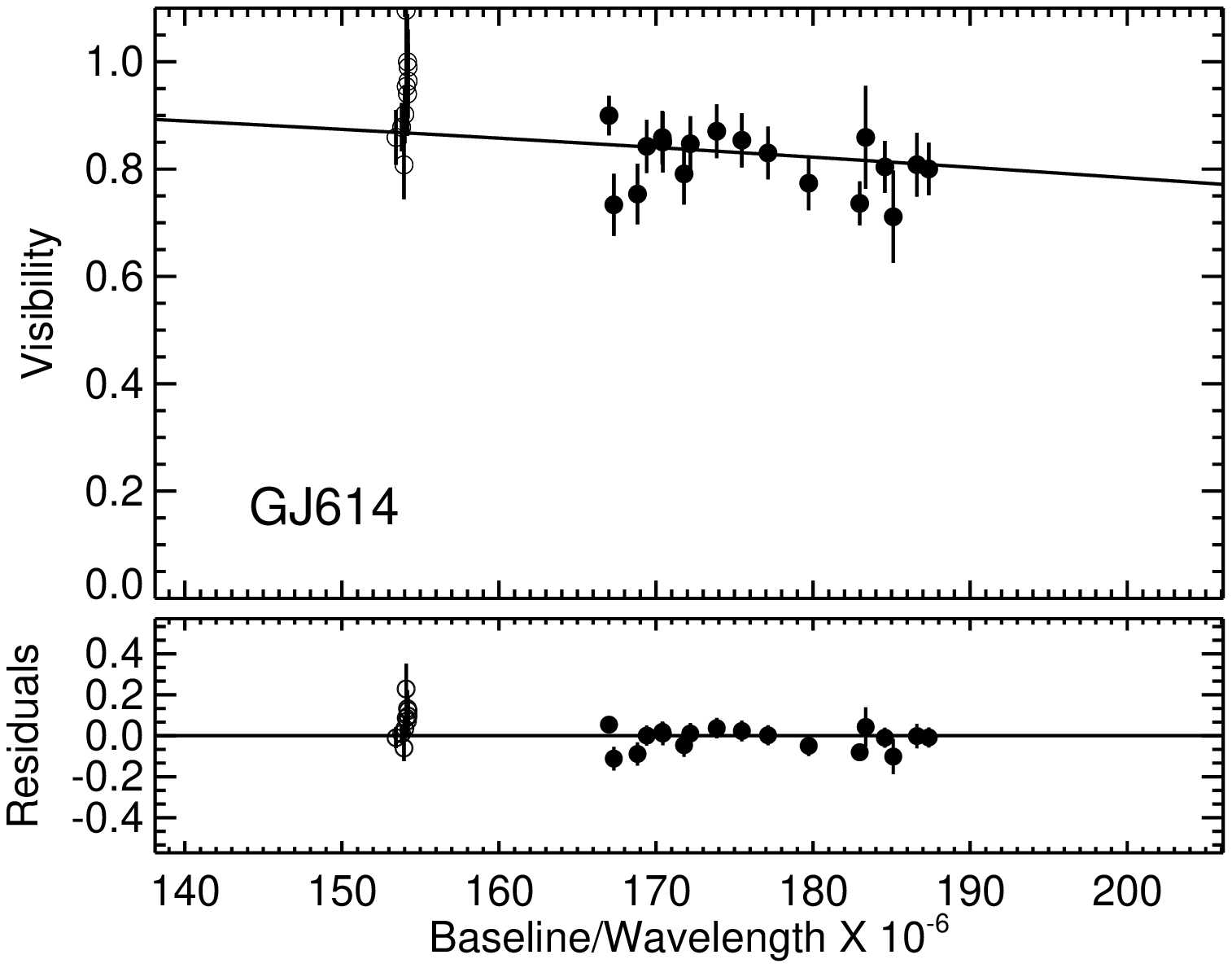} \\
      \includegraphics[angle=0,width=8.2cm]{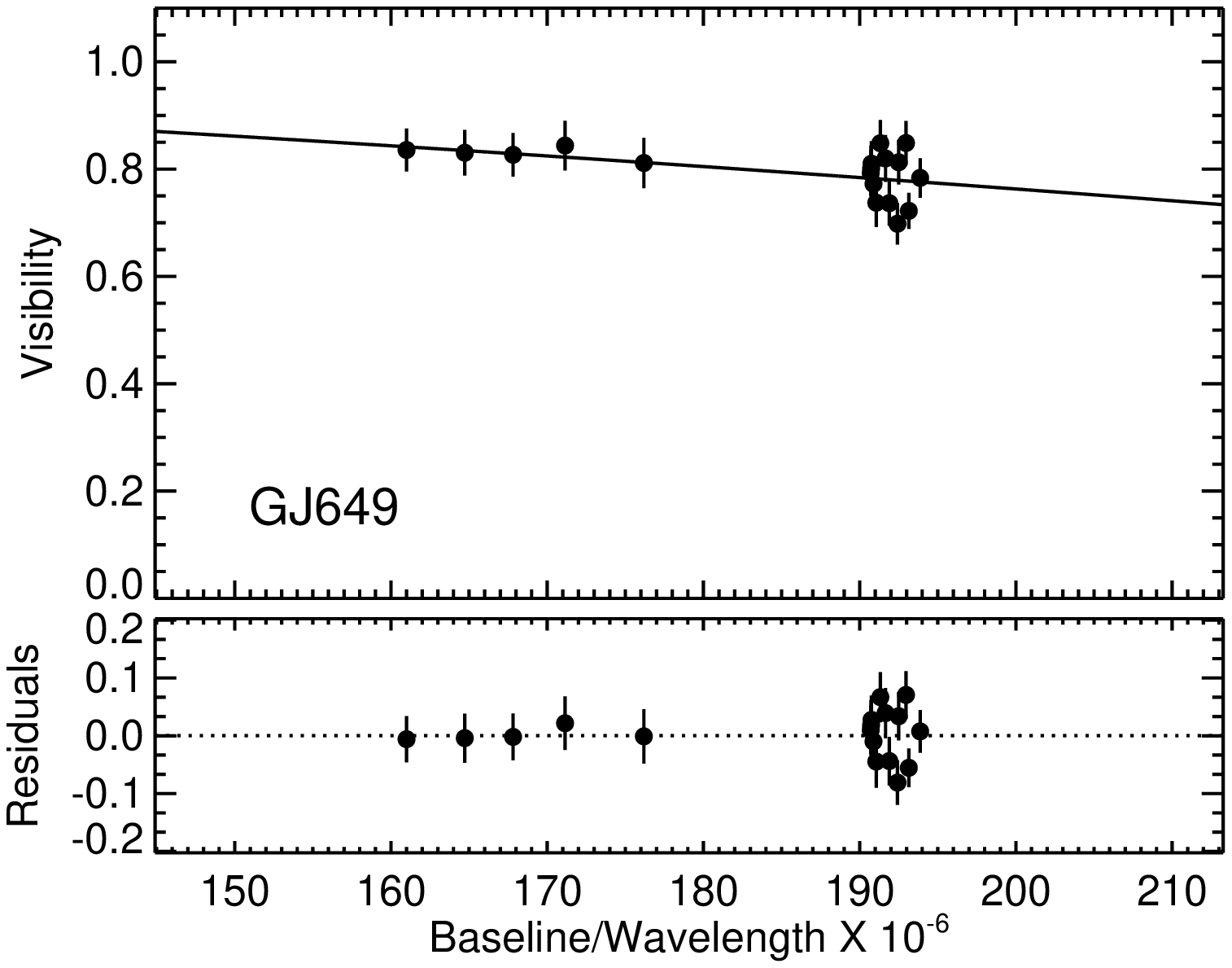} &
      \includegraphics[angle=0,width=8.2cm]{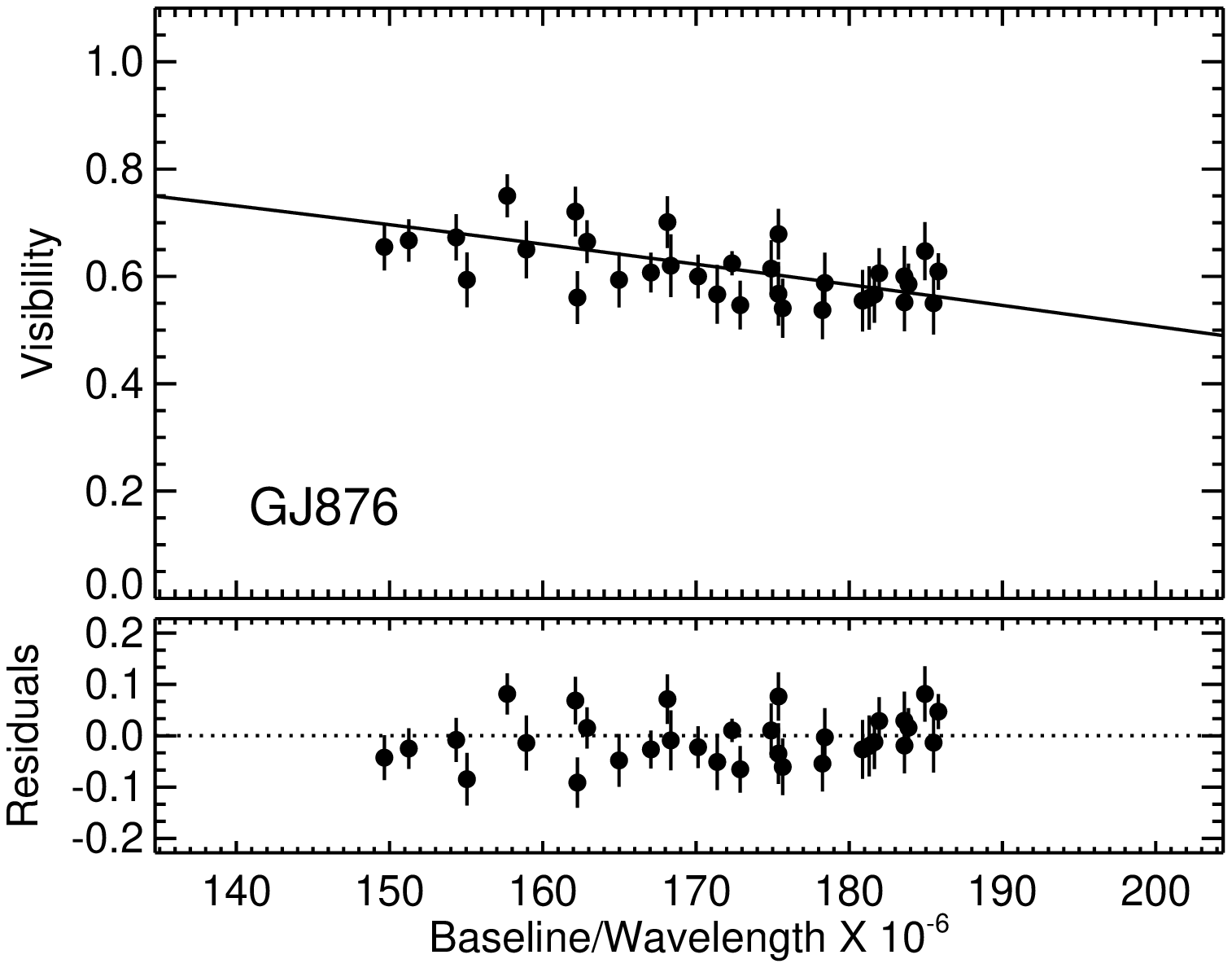} \\               
     \end{tabular}
  \end{center}
  \caption{Plots of calibrated interferometric visibilities and fits. The separated cluster of data points shown as open circles in the visibility fit for GJ~614 with a relatively large spread in visibility that is located at smaller spatial frequencies (lower numeric values in baseline / wavelength) is comprised of literature $K'$ data from \citet{bai08} -- see \S \ref{sec:gj614} for details. The interferometric observations are described in \S \ref{sec:observations}.}
  \label{fig:visibilities1}
\end{figure*}

An additional measure to reduce the influence of systematics is the alternating between multiple interferometric calibrators during observations to eliminate effects of atmospheric and instrumental systematics. Calibrators, whose angular sizes are estimated using size estimates from the Jean-Marie Mariotti Center JMDC Catalog at http://www.jmmc.fr/searchcal \citep{bon06,bon11,laf10a,laf10b}, are chosen to be small sources of similar brightness as, and small angular distance to, the respective target. A log of the interferometric observations can be found in Table \ref{tab:observations}.\footnote{As we show in Fig. \ref{fig:visibilities1} and Tables \ref{tab:observations} and \ref{tab:angular_diameters}, our angular diameter fit for GJ~614 contains literature $K'$ data obtained in 2006 and published in \citet{bai08}. }

\begin{figure*}
  \begin{center}
    \begin{tabular}{cc}  
      \includegraphics[angle=0,width=8.2cm]{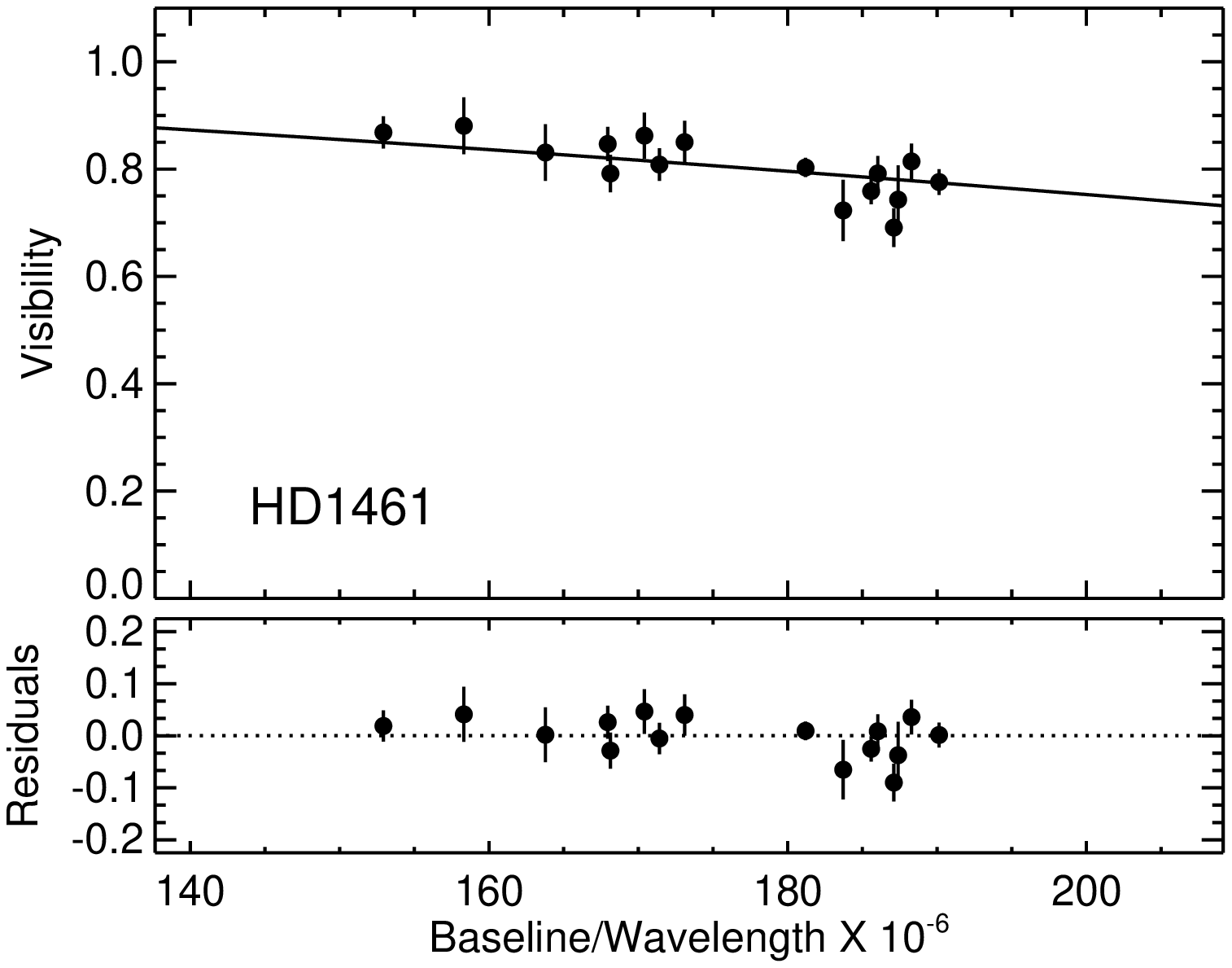} &
      \includegraphics[angle=0,width=8.2cm]{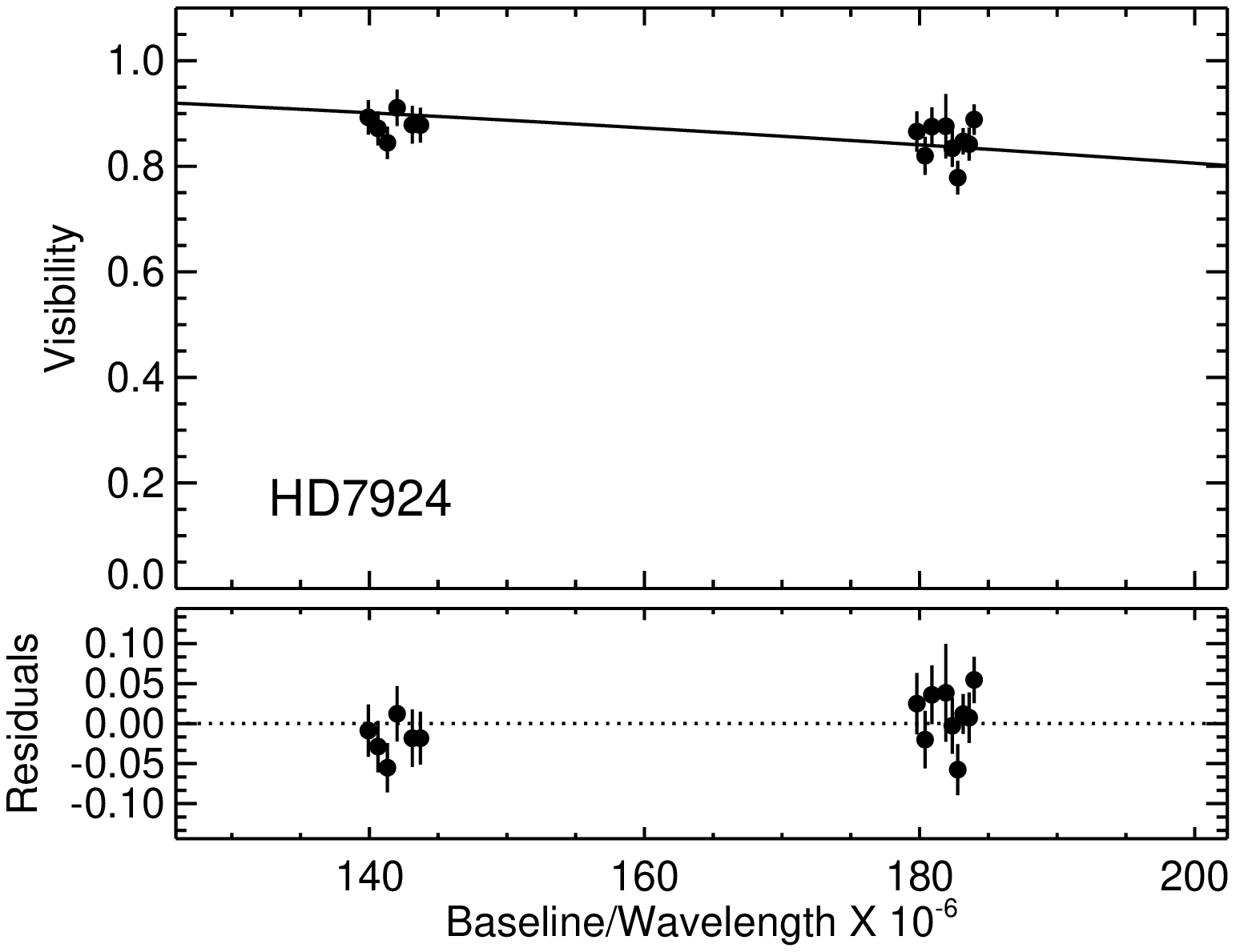} \\      
      \includegraphics[angle=0,width=8.2cm]{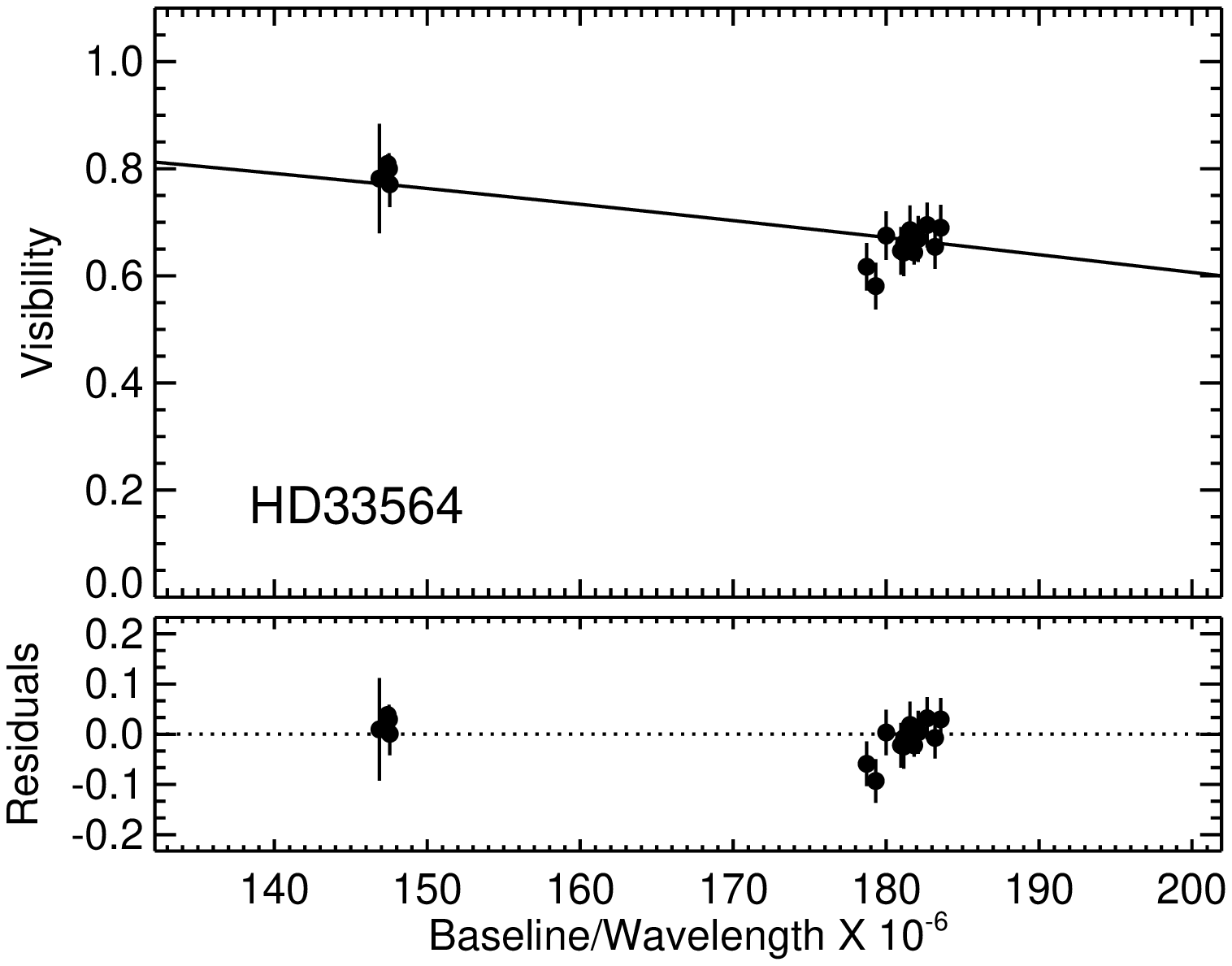} &
      \includegraphics[angle=0,width=8.2cm]{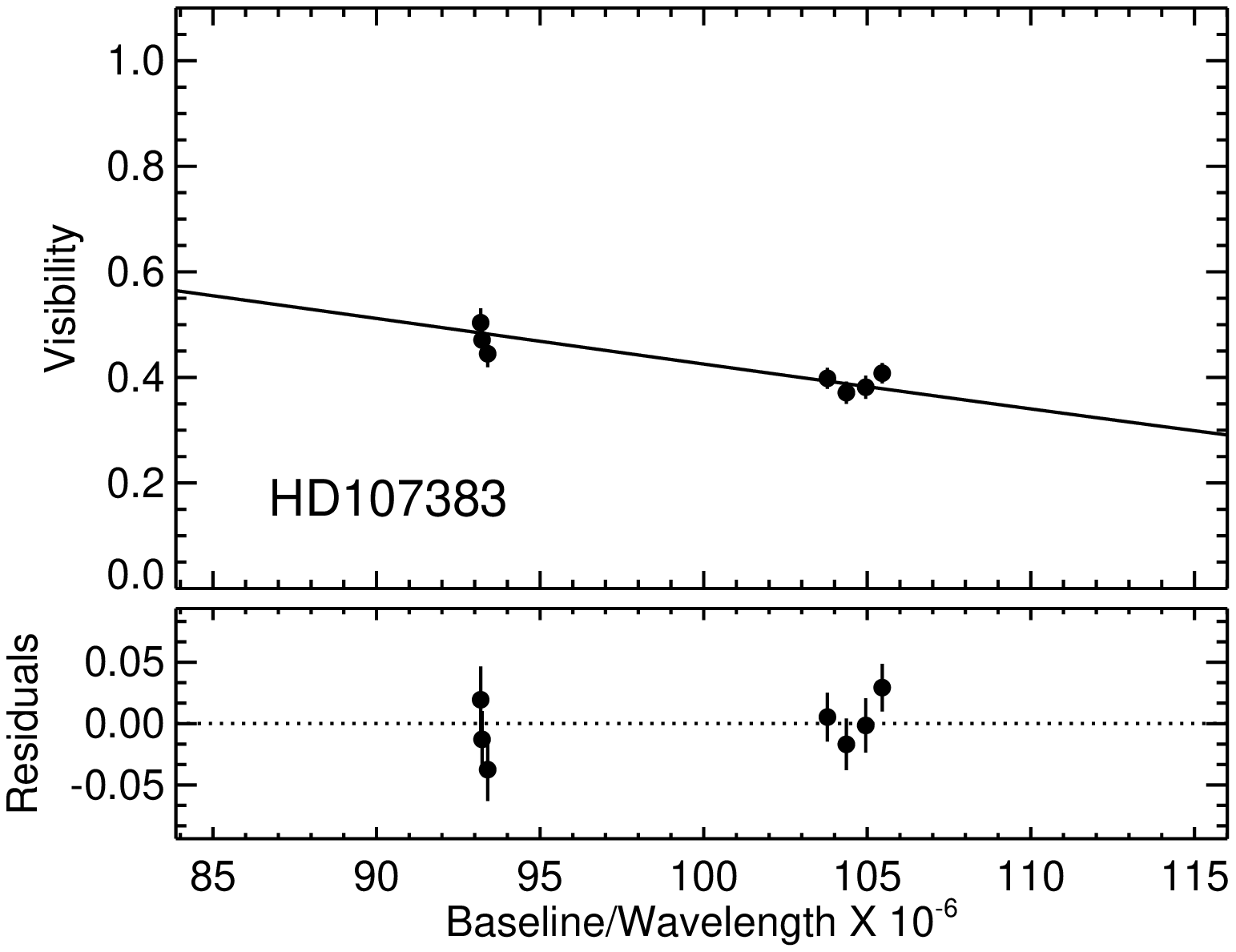} \\ 
     \end{tabular}
       \includegraphics[angle=0,width=8.2cm]{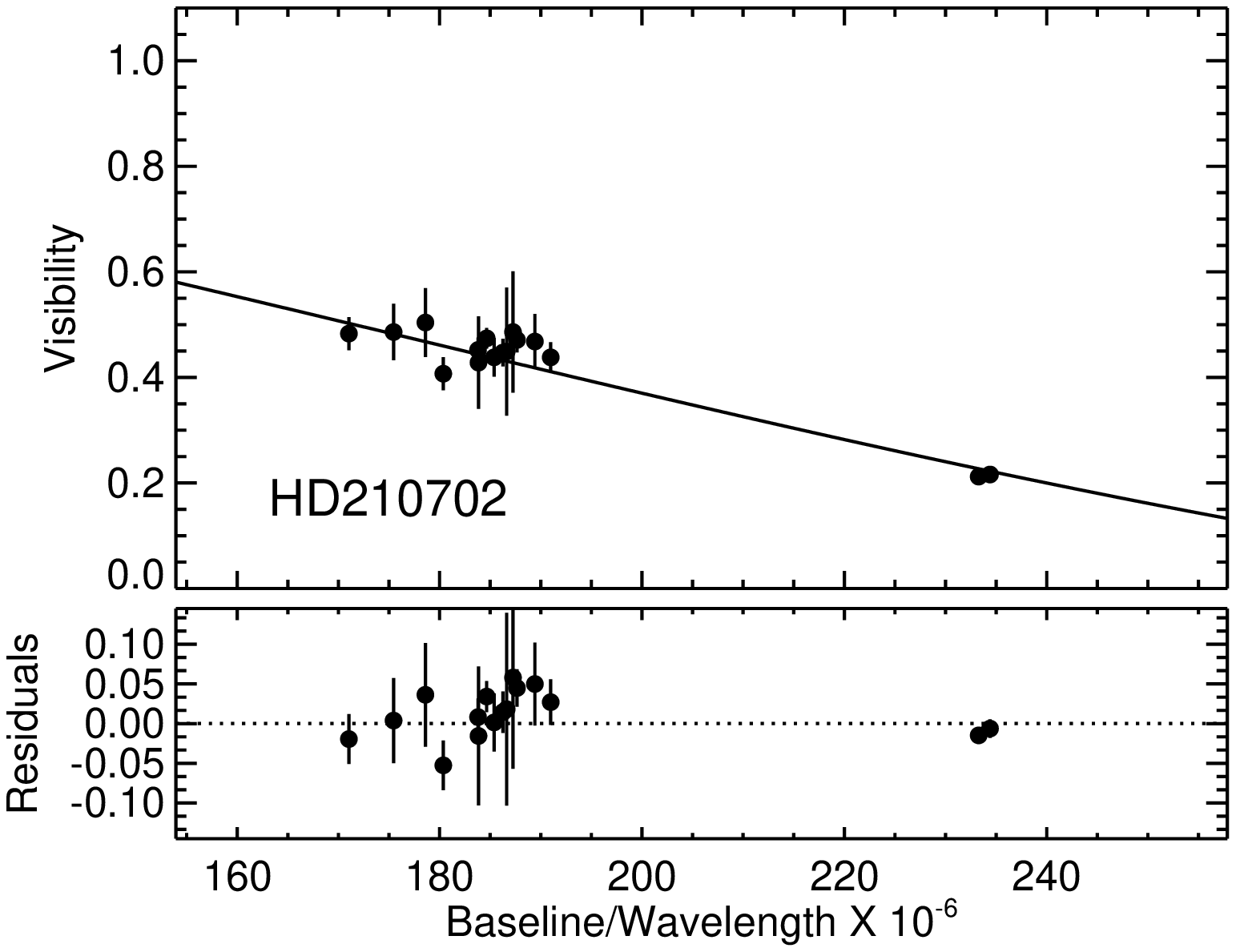}           
  \end{center}
  \caption{Plots of calibrated interferometric visibilities and fits. See \S \ref{sec:observations} for
    details on the interferometric observations.}
  \label{fig:visibilities2}
\end{figure*}

The uniform disk and limb-darkened angular diameters ($\theta_{\rm UD}$ and $\theta_{\rm LD}$, respectively; see Table \ref{tab:angular_diameters}) are found by fitting the calibrated visibility measurements (Fig. \ref{fig:visibilities1} and \ref{fig:visibilities2}) to the respective functions for each relation\footnote{Calibrated visibility data are available on request.}.  These functions may be described as $n^{th}$-order Bessel functions that are dependent on the angular diameter of the star, the projected distance between the two telescopes and the wavelength of observation \citep{han74}\footnote{Visibility is the normalized amplitude of the correlation of the light from two telescopes. It is a unitless number ranging from 0 to 1, where 0 implies no correlation, and 1 implies perfect correlation. An unresolved source would have perfect correlation of 1.0 independent of the distance between the telescopes (baseline). A resolved object will show a decrease in visibility with increasing baseline length. The shape of the visibility versus baseline is a function of the topology of the observed object (the Fourier Transform of the object's brightness distribution in the observed wavelength band). For a uniform disk this function is a Bessel function, and for this paper, we use a simple model of a limb darkened variation of a uniform disk.}. The temperature-dependent limb-darkening coefficients, $\mu_\lambda$, used to convert from $\theta_{\rm UD}$ to $\theta_{\rm LD}$, are taken from \citet{cla00} after we iterate based on the effective temperature value obtained from initial spectral energy distribution fitting (see \S \ref{sec:fbol}). Limb-darkening coefficients are dependent on assumed stellar effective temperature, surface gravity, and weakly on metallicity. When we vary the input $T_{\rm eff}$ by 200 K and $\log g$ by 0.5 dex, the resulting variations are below 0.1\% in $\theta_{\rm LD}$ and below 0.05\% in $T_{\rm eff}$. Varying the assumed metallicity across the range of our target sample does not influence our final values of $\theta_{\rm LD}$ and $T_{\rm eff}$ at all.

The values for $\theta_{\rm UD}$ and $\theta_{\rm LD}$ for our targets are given in Table~\ref{tab:angular_diameters}. The angular diameters and their respective uncertainties are computed using MPFIT, a non-linear least-squares fitting routine in IDL \citep{mar09}. Table \ref{tab:angular_diameters} shows the empirical $\chi^{2}_{reduced}$ values of the fits shown in Figures \ref{fig:visibilities1} and \ref{fig:visibilities2} in column 3. These $\chi^{2}_{reduced}$ values are often calculated to be $<<1$ due to the difficulty of accurately defining uncertainties in the visibility measurements\footnote{While there are methods of tracking errors through the calibration of visibility via standard statistical methods \citep[e.g.,][]{van05}, the principal difficulty in assessing a realistic estimate of the absolute error in visibility is the constantly changing nature of the atmosphere.}.
Consequently, we assume a true $\chi^{2}_{reduced}$ = 1 when calculating the uncertainties for $\theta_{\rm UD}$ and $\theta_{\rm LD}$, based on a rescaling of the associated uncertainties in the visibility data points. That is, the estimates of our uncertainties in $\theta_{\rm UD}$ and $\theta_{\rm LD}$ are based on a $\chi^{2}_{reduced}$ fit, not on strictly analytical calculations.


\begin{table}
\centering
\caption{Log of Interferometric Observations\label{tab:observations}}
\begin{tabular}{@{}rccc@{}}
\hline
\textbf{Star} &
 &
\# of &
 \\
 UT Date &
 Baseline &
 Obs (filter) &
 Calibrators \\
\hline
\\

\textbf{61~Vir} \\
2012/04/09 & W1/E1 & 13($H$) & HD~113289, HD~116928 \\ 
2012/04/10 & S1/E1 & 3($H$) & HD~113289, HD~116928 \\	

\\
\textbf{$\rho$~CrB} \\
2013/05/03 & S1/E1 & 4($H$)2($K^{\prime}$) & HD~139389, HD~149890 \\	
2013/08/18	&	S1/W1	& 3($H$)	&	HD~139389, HD~146946 \\	

\\
\textbf{GJ~176} \\	
2010/09/16 & W1/E1 & 7($H$) & HD~29225, HD~27524	\\	
2010/09/17 & W1/E1  & 2($H$) & HD~27534 \\	
2010/09/20 & S1/E1 & 4($H$) &  HD~29225	\\	
2010/11/10 & W1/E1 & 3($H$) & HD~29225	\\ 

\\

\textbf{GJ~614}\tablenotemark{a} \\
2010/06/28 & W1/E1 & 10($H$) & HD~144579, HD~142908 \\	
2010/06/29 & W1/E1 & 6($H$) & HD~144579, HD~142908 \\ 
2010/09/18 & S1/E1 & 4($H$) & HD~144579	\\	

\\
\textbf{GJ~649} \\
2010/06/29 & W1/E1 & 5($H$) & HD~153897 \\ 
2010/06/30 & S1/E1 & 6($H$) & HD~153897, HD150205 \\ 
2010/07/01 & S1/E1 & 7($H$) & HD~153897, HD150205 \\ 
\\

\textbf{GJ~876} \\
2011/08/17 & S1/E1 & 11($H$) & HD~215874, HD~217681 \\ 
2011/08/18 & S1/E1 & 10($H$) & HD~215874, HD~217681 \\ 
2011/08/19 & W1/E1 & 6($H$) & HD~215874, HD~216402 \\ 
2011/08/20 & W1/E1 & 6($H$) & HD~217861, HD~216402 \\ 
\\
\textbf{HD~1461} \\
2011/08/22 & S1/E1	& 7($H$)  & HD~966, HD~1100\\	
2011/10/03 & S1/E1	&	7($H$) & HD~966, HD~1100\\ 
2013/08/17	&	E1/W1	&	5($H$)	& HD~966, HD~1100\\ 

\\
\textbf{HD~7924} \\
2010/09/17 & W1/E1  & 9($H$) & HD~9407, HD~6798 \\  
2011/08/21 & W1/S1 & 6($H$) & HD~9407, HD~6798  \\ 

\\
\textbf{HD~33564} \\
2010/09/16 & W1/E1 & 2($H$) & HD~29329 \\ 
2010/09/17 & W1/E1 & 1($H$) & HD~62613 \\ 
2010/11/10 & W1/E1 & 9($H$) & HD~36768, HD~46588 \\ 
2013/08/16	&	S1/W1	&	3($H$)	&	HD~29329, HD~62613	\\ 
2013/08/17	&	S1/W1	&	1($H$)	&	HD~29329, HD~36768, HD~46588	\\ 

\\
\textbf{HD~107383} \\
2013/05/06 & E2/W2 & 3($H$) & HD~106661, HD~108468 \\ 
2013/05/06 & S2/W2 & 4($H$) & HD~106661, HD~104452  \\ 

\\
\textbf{HD~210702} \\
2013/08/18 & S1/E1 & 5($H$)1($J$) & HD~210074, HD~206043 \\ 
2013/08/19 & E1/W1  & 2($H$) & HD~210074, HD~206043  \\ 
2013/08/22 & E1/W1  & 7($H$) & HD~210074, HD~207223  \\ 
2013/08/22 & S1/E1  & 1($J$) & HD~210074, HD~207223  \\ 

\\
\tablenotetext{a}{We combine our data with literature CHARA $K'$-band data points, observed in 2006 with the S1/E1 baseline, published in \citet{bai08} -- see \S \ref{sec:gj614} for additional details.} 
\tablecomments{For details on the interferometric observations, see \S\ref{sec:observations}.} 
\end{tabular}
\end{table}


\begin{table*}
\centering
\caption{Stellar Angular Diameters\label{tab:angular_diameters}}
\begin{tabular}{@{}lcccccc@{}}
\hline
Star &
\# of &
Reduced &
$\theta_{\rm UD} \pm \sigma$ &
 &
$\theta_{\rm LD} \pm \sigma$ &
$\theta_{\rm LD}$ \\
Name &
Obs.	&
$\chi^{2}$	&
(mas) &
$\mu_{\lambda}$	&
(mas)	&
\% err \\
\hline
61~Vir	&	16	&	0.31	&	$1.037\pm0.005$	&	0.367	&	$1.073\pm0.005$	&	0.4	\\
$\rho$~CrB	&	9 &       0.46 & $0.714\pm0.013$ &      0.342 & $0.735\pm0.014$ &        1.9 \\ 	
GJ~176	&	14	&	0.18	&	$0.441\pm0.020$	&	0.210	&	$0.448\pm0.021$	&	4.6	\\
GJ~614 &     28\tablenotemark{a} &        1.16 & $0.449\pm0.017$ &      0.284 & $0.459\pm0.017$ &        3.7 \\  
GJ~649	&	18	&	1.02	&	$0.472\pm0.012$	&	0.327	&	$0.484\pm0.012$	&	2.5	\\
GJ~876	&	33	&	0.32	&	$0.721\pm0.009$	&	0.398	&	$0.746\pm0.009$	&	1.2	\\
 HD~1461 &     16 &       0.19 & $0.483\pm0.010$ &      0.369 & $0.498\pm0.011$ &        2.1 \\
HD~7924	&	15	&	0.19	&	$0.424\pm0.014$	&	0.281	&	$0.433\pm0.014$	&	3.2	\\	
 HD~33564 &     16 &       1.08 & $0.629\pm0.010$ &      0.225 & $0.640\pm0.010$ &        1.6 \\
HD~107383	&	7	&	0.16	&	$1.590\pm0.015$	&	0.417	&	$1.651\pm0.016$	&	1.0	\\
 HD~210702 &     16 &       1.57 & $0.845\pm0.005$ &      0.484 & $0.886\pm0.006$ &        0.6 \\
\hline
\end{tabular}
\tablenotetext{a}{Includes CHARA Classic $K^{\prime}$ data from \citet{bai08}.}
\tablecomments{$\theta_{UD}$ and $\theta_{LD}$ refer to stellar uniform-disk and limb-darkening-corrected angular diameters, respectively. $\mu_\lambda$ are the limb-darkening coefficients from \citet{cla00} after an iteration based on $T_{\rm eff}$ values. Refer to \S \ref{sec:observations} for details.}

\end{table*}


\subsection{Bolometric Fluxes} \label{sec:fbol}

In this Section, we report on stellar spectral energy distributions (SEDs) fits
of our targets. We augment literature broad-band photometry data by using spectrophotometric data whenever available. The purpose of these SED fits is to obtain direct estimates of stellar $T_{\rm eff}$ and $L$, as described in \S \ref{sec:direct}. Our procedure is analogous to the ones performed in \citet{van09,von11c,von11a,von12,boy12a,boy12b,boy13}.

\begin{figure*}										%
\centering
\epsfig{file=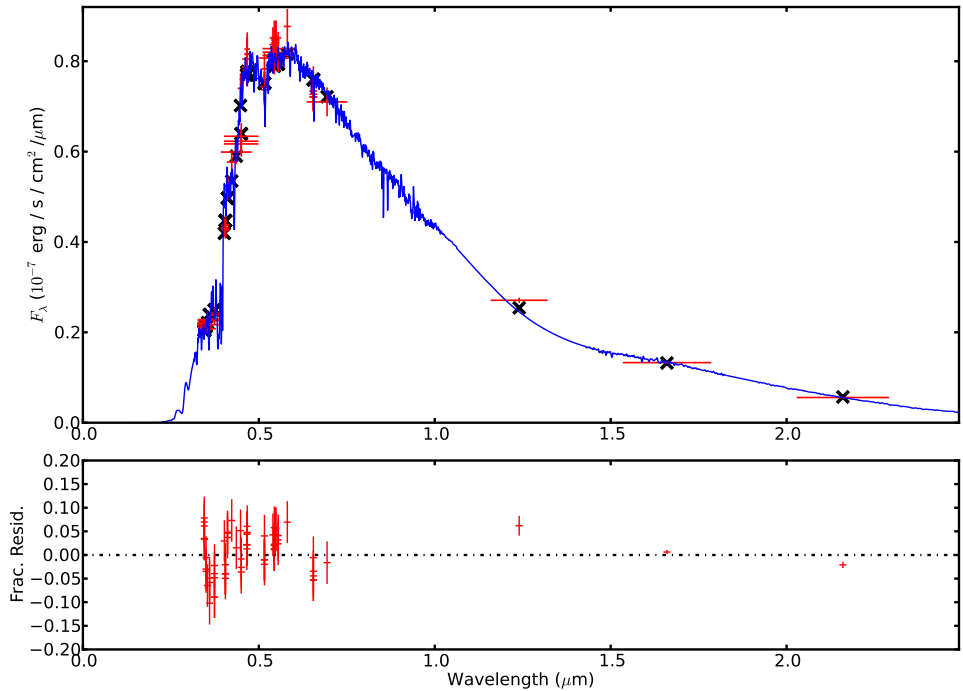,angle=0, width=15cm} \\
\caption{SED Fit of GJ~614 to illustrate our fitting routine. In the top panel, the (blue) spectrum is a G9 IV spectral template from the \citet{pic98} library. The (red) crosses indicate photometry values from the literature. ``Error bars" in x-direction represent the bandwidths of the photometric filters. The (black) X-shaped symbols indicate the flux value of the spectral template averaged over the filter transmission range in wavelength. The lower panel shows the residuals around the fit in fractional flux units of photometric uncertainty. The uncertainties in y-direction in the lower plot represent the photometric uncertainties in the literature data scaled by the corresponding flux values. For more details, see \S \ref{sec:fbol}.}
\label{fig:sed}
\end{figure*}

Our SED fitting is based on a $\chi^2$-minimization of input SED templates from the \citet{pic98} to literature photometry of the star under investigation. If the literature photometry values are in magnitudes, they are converted to absolute fluxes by application of published or calculated zero points. The filters of the literature photometry data are assumed to have a top-hat shape. That is, during the calculation of $\chi^2$, only the central filter wavelengths are correlated with the SED template's flux value averaged over the filter transmission range in wavelength. Literature spectrophometry data are very useful for SED fitting since multiple individual photometry data points, instead of being integrated into a single wavelength, trace out the shape of the SED in great detail. The SED template is scaled to minimize $\chi^2$ and then integrated over wavelength to obtain the bolometric flux. The code additionally produces an estimated angular diameter, which we only use as a sanity check to avoid systematic problems like the choice of wrong spectral template. Figure \ref{fig:sed} illustrates our procedure for the example of GJ~614.

We note that our quoted uncertainties on the bolometric flux values are statistical only. We do not (and indeed cannot) account for possible systematics such as saturation or correlated errors in the photometry, filter errors due to problems with transmission curves, or other non-random error sources. The only systematics that we can control are (1) the choice of spectral template for the SED fit, (2) the choice of which photometry data to include in the fit, and (3) whether to let the interstellar reddening float during the fit or whether to set it to zero. In order to be consistent as possible, we take the following approach:

\begin{itemize}
\item All photometry data are included in the fit in principle, except when they present clear outliers in the SED. This way, we attempt to reduce systematics. In general, there are tens to hundreds of photometric measurements per target (see Table \ref{tab:fbol}), and at most, we remove 1--2 data points, mostly in the $U$ band or $RI$ bands. 
\item Interstellar extinction is set to zero for all targets, due to the small distances to the stars (see Table \ref{tab:parameters}), which are adopted from \citet{van07}. We cross-checked results with the ones using variable reddening, and in almost no case was there any difference. The ones for which the variable reddening produced results that are not consistent with $A_V = 0$ at the $\sim$1-$\sigma$ level are discussed below. For all others, letting $A_V$ vary produced $A_V=0$.
\item Whenever a literature photometry datum has no quoted uncertainty associated with it, we assign it a 5\% random uncertainty. This is only the case for some older data sets. 
\item The choice of spectral template is based on minimization of $\chi_{reduced}^2$ only. For about half of our targets, we linearly interpolate between the relatively coarse grid of the \citet{pic98} spectral templates to obtain a better fit and thus a more accurate value for the numerical integration to calculate the bolometric flux (indicated in Table \ref{tab:fbol}). Linear interpolation never spans more than three tenths in spectral type range (e.g., G5 to G8). 
\item Despite the fact that spectrophotometry often increases the fit's $\chi_{reduced}^2$, we include these data whenever available (indicated in Table \ref{tab:fbol}) in order to reduce the systematics in the choice of spectral template, which is determined more accurately via spectrophotometry.
\end{itemize}

Notes on individual systems with respect to SED fitting: 
\begin{itemize}

\item \textbf{61~Vir:} Despite the fact that we find $A_V=0$ when letting $A_V$ float, we note that dust excess for this system was reported in \citet{tri08,tan09,bry09,law09}. Photometry sources: \citet{joh66,joh75,gol72,dea81,hag87,ols94,oja96}; spectrophotometry from \citet{bur85}.
\item \textbf{$\rho$~CrB:} Photometry sources: \citet{arg63,gol72,cla79,hag87,bei88,jas90,kor91,ski94,glu98,gez99,cut03}; spectrophotometry from \citet{bur85}.
\item \textbf{GJ~176:} Photometry sources: \citet{wei84,sta86,wei86,wei87,wei93,wei96,gez99,bes00,cut03}.
\item \textbf{GJ~614:} Photometry sources: \citet{arg63,gol72,bei88,kor91,cut03,kaz05}. 
\item \textbf{GJ~649:} Photometry sources: \citet{mum56,jon81,mer86,sta86,bei88,wei93,wei96,cut03,kaz05}. 
\item \textbf{GJ~876:} With a variable $A_V$, the SED fit for GJ~876 produces slightly different results compared to the ones given in Table \ref{tab:fbol}, where it was set to zero: $\chi^2_{red}$ = 4.86, $F_{\rm BOL} = (1.95 \pm 0.001) \times 10^{-8}$ erg $\rm s^{-1} \rm cm^{-2}$ with an $A_V = 0.111 \pm 0.007$. Photometry sources: \citet{err71,iri71,gol72,mou76,per77,jon81,wei82,the84,koz85,wei86,wei87,bes90a,wei96,bes00,koe02,cut03,kil07,koe10}. 
\item \textbf{HD~1461:} Photometry sources: \citet{cou62,gol72,spe81,hag87,kor91,cut03,kaz05}. 
\item \textbf{HD~7924:} Photometry sources: \citet{san66,gol72,kor91,cut03}. 
\item \textbf{HD~33564:} With a variable $A_V$, the SED fit for HD~33564 produces slightly different results compared to the ones given in Table \ref{tab:fbol}: $\chi^2_{red}$ = 3.81, $F_{\rm BOL} = (25.05 \pm 0.19) \times 10^{-8}$ erg $\rm s^{-1} \rm cm^{-2}$ with an $A_V = 0.076 \pm 0.007$. Photometry sources: \citet{hag66,gol72,bei88,cer89,gez99,kor91,cut03}; spectrophotometry from \citet{kha88}.
\item \textbf{HD~107383:} Photometry sources: \citet{joh66,hag70,str70,gol72,joh75,bei88,kor91,yos97,cut03}. 
\item \textbf{HD~210702:} With a variable $A_V$, the SED fit for HD~210702 produces marginally different results compared to the ones given in Table \ref{tab:fbol}: $\chi^2_{red}$ = 0.77, $F_{\rm BOL} = (14.24 \pm 0.05) \times 10^{-8}$ erg $\rm s^{-1} \rm cm^{-2}$ with an $A_V = 0.037 \pm 0.029$. Photometry sources: \citet{joh57,joh66,gol72,ols74,mcc81,kor91,cut03}. 

\end{itemize}


\begin{table}
\caption{$F_{\rm BOL}$ Values from SED Fitting \label{tab:fbol}}
\begin{tabular}{@{}lcccc@{}}
\hline
 &
Template &
Degrees of &
 &
$F_{\rm BOL} \pm \sigma$\\
Star &
Sp. Type &
Freedom &
$\chi^2_{red}$ &
($10^{-8}$ erg $\rm s^{-1} \rm cm^{-2}$) \\
\hline
61~Vir & G6.5V\tablenotemark{a} & 233\tablenotemark{b} & 2.81 & 36.06 $\pm$ 0.05\\
$\rho$~CrB & G0V & 369\tablenotemark{b} & 8.74 & 18.03 $\pm$ 0.02\\
GJ~176 & M2.5V & 36 & 5.95 & 1.26 $\pm$ 0.005\\
GJ~614 & G9IV\tablenotemark{a} & 69 & 1.28 & 6.50 $\pm$ 0.02\\
GJ~649 & M2V & 32 & 8.90 & 1.30 $\pm$ 0.005\\
GJ~876 & M3.5V\tablenotemark{a} & 109 & 7.18 & 1.78 $\pm$ 0.004\\
HD~1461 & G2IV & 91 & 0.63 & 6.95 $\pm$ 0.03\\
HD~7924 & K0.5V\tablenotemark{a} & 16 & 2.37 & 4.14 $\pm$ 0.03\\
HD~33564 & F6V & 119\tablenotemark{b} & 4.88 & 23.17 $\pm$ 0.04\\
HD~107383 & K0.5III\tablenotemark{a} & 68 & 1.60 & 44.49 $\pm$ 0.25\\
HD~210702 & G9III\tablenotemark{a} & 47 & 0.79 & 13.65 $\pm$ 0.09\\
\hline
\end{tabular}
\tablecomments{For more details, please see \S \ref{sec:fbol}.}
\tablenotetext{a}{Linearly interpolated between \citet{pic98} spectral templates.}
\tablenotetext{b}{Includes spectrophotometry.}
\end{table}


\section{Stellar Astrophysical Parameters}\label{sec:properties}

In this Section, we report on our direct measurements of stellar diameters, $T_{\rm eff}$, and luminosities, based on our interferometric measurements (\S \ref{sec:observations}) and SED fitting to literature photometry (\S \ref{sec:fbol}). We furthermore present calculated values wherever sensible for our targets, such as the location and extent of the respective system's circumstellar habitable zone (HZ) and stellar mass and age. Results are summarized in Table \ref{tab:parameters}.


\subsection{Direct: Stellar Radii, Effective Temperatures, and Luminosities} \label{sec:direct}

We use our measured, limb-darkening corrected, angular diameters $\theta_{\rm LD}$, corresponding to the angular diameter of the Rosseland, or mean, radiating surface of the star (\S \ref{sec:observations}, Table \ref{tab:angular_diameters}), coupled with trigonometric parallax values from \citet{van07} to determine the linear stellar diameters. Uncertainties in the physical stellar radii are typically dominated by the uncertainties in the angular diameters, not the distance.

From the SED fitting (\S \ref{sec:fbol}, Table \ref{tab:fbol}), we calculate the value of the stellar bolometric flux, $F_{\rm BOL}$ by numerically integration of the scaled spectral template across all wavelengths. Wherever the empirical spectral template does not contain any data, it is interpolated and extrapolated along a blackbody curve. Combination with the rewritten version of the Stefan-Boltzmann Law

\begin{equation} \label{eq:temperature}
T_{\rm eff} ({\rm K}) = 2341 (F_{\rm BOL}/\theta_{\rm LD}^2)^{\frac{1}{4}},
\end{equation}

\noindent
where $F_{\rm BOL}$ is in units of $10^{-8}$~erg cm$^{-2}$ s$^{-1}$ and $\theta_{\rm LD}$ is in units of mas, produces the effective stellar temperatures $T_{\rm eff}$. 


\subsection{Calculated: Habitable Zones, Stellar Masses and Ages}\label{sec:calculated}

To first order, the habitable zone of a planetary system is described as the range of distances in which a planet with a surface and an atmosphere containing a modest amount of greenhouse gases would be able to host liquid water on its surface, first characterized in \citet{kas93}. The HZ boundaries in this paper are calculated based on the updated formalism of \citet{kop13a,kop13b}\footnote{We use the online calculator at http://www3.geosc.psu.edu/$\sim$ruk15/planets/.}. \citet{kop13b} define the boundaries based on a runaway greenhouse effect or a runaway snowball effect as a function of stellar luminosity and effective temperature, plus water absorption by the planetary atmosphere. Whichever assumption is made of how long Venus and Mars were able to retain liquid water on their respective surfaces defines the choice of HZ (conservative or optimistic). These conditions are described in more detail in \citet{kop13b} and section 3 of \citet{kan13}. The boundaries quoted in Table \ref{tab:parameters} are the ones for both the conservative and optimistic HZ. 

For stellar mass and age estimates of the early stars in our sample, we use the Yonsei-Yale ($Y^2$) isochrones \citep{yi2001,kim02,dem04}. Input data are our directly determined stellar radii and temperatures, along with the literature metallicity values from Table~\ref{tab:parameters}, and zero $\alpha$-element enhancement: [$\alpha$/Fe]~$=0$. We follow the arguments in section 2.4 in \citet{boy13} and conservatively estimate mass and age uncertainties of 5\% and 5~Gyr, respectively. 

The ages of low-mass stars, however, are not sensitive to model isochrone fitting. Thus, in order to estimate the stellar masses of the KM dwarfs in our sample, we use the formalism described in \S 5.4 in \citet{boy12b}. We use the data from Table 6 in \citet{boy12b} to derive the following equation for KM dwarfs:  

\begin{eqnarray}
\label{eq:massradius}
	M 	&	=	&	-0.0460 (\pm 0.0251) + 1.0930 (\pm 0.1481) R + \nonumber \\
		&		&	0.0064 (\pm 0.1722) R^2,
\end{eqnarray}


\noindent where $R$ and $M$ are the stellar radius and mass in Solar units, respectively. This essentially represents the inverted form of equation 10 in \citet{boy12b} for single dwarf stars. We note that the statistical errors in the determined masses using Equation \ref{eq:massradius} are dominated by the uncertainties in the coefficients and are of order $\sim 30$\%.


\begin{table*}
\caption{Stellar Astrophysical Parameters \label{tab:parameters}}
\begin{tabular}{@{}lccccccccccc@{}}
\hline
 		&	Radius 				&	$T_{\rm eff}$	&	$L$					&	Spectral	&	Distance	&	Metallicity		&	$\rm HZ_{cons}$	&	$\rm HZ_{opt}$ &	Age	&	Mass 				\\
Star 	&	($R_{\odot}$	)	&	(K)					&	($L_{\odot}$	)	&	Type			&	(pc)			&	[Fe/H]			&	(AU)	& (AU) &	(Gyr)	&	($M_{\odot}$)	\\
\hline
61~Vir & $0.9867\pm0.0048$ & $5538\pm13$ & $0.8222\pm0.0033$ & G7 V & 8.56  & 0.01 & 0.91 -- 1.56 & 0.69 -- 1.63 & 8.6 & 0.93 \\
$\rho$~CrB & $1.3617\pm0.0262$ & $5627\pm54$ & $1.7059\pm0.0423$ & G0 V & 17.43 & -0.22 & 1.30 -- 2.23 & 0.98 -- 2.33 & 12.9 & 0.91 \\
GJ~176 & $0.4525\pm0.0221$ & $3679\pm77$ & $0.0337\pm0.0018$ & M2 & 9.27 & 0.15\tablenotemark{a} & 0.20 -- 0.37 & 0.15 -- 0.38 & -- & 0.45\tablenotemark{b} \\
GJ~614 & $0.8668\pm0.0324$ & $5518\pm102$ & $0.6256\pm0.0077$ & K0 IV-V & 15.57 & 0.44 & 0.79 -- 1.37 & 0.60 -- 1.43 & -- & 0.91\tablenotemark{b} \\
GJ~649 & $0.5387\pm0.0157$ & $3590\pm45$ & $0.0432\pm0.0013$ & M0.5 & 10.34 & -0.04\tablenotemark{a} & 0.22 -- 0.42 & 0.17 -- 0.44 & -- & 0.54\tablenotemark{b} \\
GJ~876 & $0.3761\pm0.0059$ & $3129\pm19$ & $0.0122\pm0.0002$ & M4 & 4.69 & 0.19\tablenotemark{a} & 0.12 -- 0.23 & 0.09 -- 0.24 & -- & 0.37\tablenotemark{b} \\
HD~1461 & $1.2441\pm0.0305$ & $5386\pm60$ & $1.1893\pm0.0476$ & G3 V & 23.44 & 0.16 & 1.10 -- 1.89 & 0.83 -- 1.98 & 13.8 & 0.94 \\
HD~7924 & $0.7821\pm0.0258$ & $5075\pm83$ & $0.3648\pm0.0077$ & K0 V & 16.82 & -0.14 & 0.62 -- 1.08 & 0.47 -- 1.13 & -- & 0.81\tablenotemark{b}  \\
HD~33564 & $1.4367\pm0.0238$ & $6420\pm50$ & $3.1777\pm0.0696$ & F7 V & 20.98 & 0.08 & 1.73 -- 2.88 & 1.31 -- 3.00 & 2.2 & 1.31 \\
HD~107383 & $15.781\pm0.3444$ & $4705\pm24$ & $109.51\pm4.3256$ & K0 III & 88.89 & -0.30 & 10.8 -- 19.4 & 8.19 -- 20.3 & --\tablenotemark{c} & --\tablenotemark{c} \\
HD~210702 & $5.2314\pm0.1171$ & $4780\pm18$ & $12.838\pm0.5569$ & K1 III & 54.95 & 0.03\tablenotemark{d} & 3.70 -- 6.59 & 2.80 -- 6.90 & 5.0 & 1.29 \\

\hline
\end{tabular}
\tablecomments{For the calculations of stellar $R$, $T_{\rm eff}$, and $L$, please see \S \ref{sec:direct}. Spectral types and metallicities from \citet{and11,and12} unless otherwise indicated. Distances from \citet{van07}. The calculations for the inner and outer boundaries of the system circumstellar HZs (conservative and optimistic) and for stellar mass and age are described in \S \ref{sec:calculated}. }
\tablenotetext{a}{From \citet{roj12}. Each value has a quoted uncertainty of 0.17 dex.}
\tablenotetext{b}{Masses calculated via Equation \ref{eq:massradius}, based on data in \citet{boy12b}.}
\tablenotetext{c}{Beyond the range of the $Y^2$ isochrones. }  
\tablenotetext{d}{From \citet{mal13}.}
\end{table*}



\section{Discussion} \label{sec:discussion}

In this Section, we briefly discuss our results on the individual systems. We compare our directly determined stellar parameters (\S \ref{sec:direct} and Table \ref{tab:parameters}) to quoted literature values. Statistical differences between values are calculated by adding our uncertainties and those from the literature, when available, in quadrature.





\subsection{61~Vir (= HD~115617)} \label{sec:61vir}

61~Vir hosts three planets with periods ranging from 4.2 to 124 days and minimum masses between 5.1 and 24 $M_{\earth}$ \citep{vog10b}. Since the orbital distances of all three planets are less than 0.5 AU from the parent star, and since the inner edge of the optimistic/conservative HZ is located at 0.69 AU / 0.91 AU, none of the known three planets reside within the system's HZ. 
 
The stellar radius for 61~Vir quoted in \citet[][$0.98\pm0.03 R_{\odot}$]{tak07a} is statistically identical with ours.
When compared to \citet{val05}, our radius measurement is $\sim 2 \sigma$ larger than their estimate ($0.963\pm0.011 R_{\odot}$), while our temperature value is consistent with their value ($5571\pm44$ K).  Similarly good agreement exists with the $T_{\rm eff}$ value quoted in \citet[][$5577\pm33$ K]{ecu06}.

To estimate 61~Vir's mass and age, we use the $Y^2$ isochrones with our values from Table \ref{tab:parameters} as input, generated with a $0.1$~Gyr step size, and a fixed metallicity of [Fe/H]~$= +0.01$ (Table~\ref{tab:parameters}).  Interpolation within the best fitting isochrone gives an age of $8.6$~Gyr and mass of $0.93 M_{\odot}$. 


\subsection{$\rho$~CrB (= HD~143761)} \label{sec:rhocrb}

$\rho$~CrB has a Jupiter-mass planet in a 40-day orbit \citep{noy97} whose orbital semi-major axis \citep[0.22 AU;][]{but06} is located well inside the system's optimistic/conservative HZ's inner boundary at 0.98 AU / 1.3 AU. 

$\rho$~CrB's radius was interferometrically observed in $K'$-band by \citet{bai08} who calculate a radius of $1.284\pm0.082 R_{\odot}$, well within $1\sigma$ of our value of $1.3424 R_{\odot}$. $\rho$~CrB's stellar radius from \citet{fuh98}, $1.34\pm0.05 R_{\odot}$, also agrees very well ($<< 1 \sigma$) with our directly determined value. The spectroscopically determined $T_{\rm eff}$ values of \citet[][$5821\pm80$ K]{fuh98} and \citet[][$5822\pm44$ K]{val05}, however, come in at $1.8\sigma$ and $2.7\sigma$ above our estimate of 5665 K.  

Using the same procedure as for 61~Vir (\S \ref{sec:61vir}), we solve for best-fit $Y^2$ isochrone properties for $\rho$~CrB using a metallicity of [Fe/H]~$= -0.22$ (Table~\ref{tab:parameters}).  This produces estimates for stellar age of 12.9~Gyr and stellar mass of $0.91 M_{\odot}$.  



\subsection{GJ~176 (= HD~285968)} \label{sec:gj176}

The 8.8 Earth-mass planet in a 10.24 day, circular orbit around GJ~176 \citep{end08,for09} is at a projected semi-major axis of less than 0.07 AU from its parent star, well inside the inner boundary of the system optimistic/conservative HZ at 0.15 AU / 0.2 AU. 






\citet{tak07} estimate GJ~176's stellar radius to be $0.46 ^{+ 0.01}_{-0.02} R_{\odot}$, statistically identical to our directly determined value of $0.4525 R_{\odot}$. The $T_{\rm eff}$ estimate in \citet[][3520~K]{mor08} is around $2\sigma$ below our value, in part due to the fact, however, that no uncertainties in $T_{\rm eff}$ are provided. Finally, our values are in agreement with $T_{\rm eff} = 3754$~K and $R=0.40 R_{\odot}$ from \citet{wri11a}. 

We estimate the mass of GJ 176 to be $0.45 M_{\odot}$ via Equation~\ref{eq:massradius}. It is possible, but extremely unreliable, to estimate isochronal ages for low-mass stars, as stated earlier in \S \ref{sec:calculated}.





\subsection{GJ~614 (= HD~145675 = 14 Her)} \label{sec:gj614}

The planet in orbit around GJ~614 was first announced by M. Mayor in 1998 (see http://obswww.unige.ch/$\sim$udry/planet/14her.html). \citet{but03} provide system characterization, including the period of around 4.7 years with an eccentricity of 0.37 and a planet minimum mass of about 4.9 Jupiter masses. At a projected semi-major axis of 2.82 AU, this planet is actually located outside the system optimistic/conservative HZ's outer edge at 1.43 AU / 1.37 AU, even at periastron distance $r_{peri} = (1-e) a = 1.78$ AU.

\citet{bai08} interferometrically determine GJ~614's physical radius to be $0.708\pm0.85 R_{\odot}$, which is around 1.9$\sigma$ below our value. Those data are taken in the $K'$ band, where GJ~614's small angular diameter is only marginally resolved. Since the \citet{bai08} $K^{\prime}$ band data alone thus do not constrain the angular diameter fit as well as our $H$ data, we combine the \citet{bai08} visibilities along with our own to determine GJ~614's angular diameter in Table~\ref{tab:angular_diameters} and linear radius in Table~\ref{tab:parameters}. Our $H$-band data enable increased resolution due to the shorter wavelengths (see Figure~\ref{fig:visibilities1}, where the \citealt{bai08} calibrated visibilities are represented by the cluster of data points with relatively large scatter at shorter effective baseline values).

There are a large number (many tens on Vizier) of $T_{\rm eff}$ values in the literature with estimates between 4965~K and 5735~K, which average slightly over 5300~K. The catalog in \citet{sou10a} alone contains 19 values between 5129~K \citep{ram05} and 5600~K \citep{hei03}. Our value of $5518 \pm 102$~K falls toward the upper tier of the literature range. 


Using Equation~\ref{eq:massradius}, we derive the stellar mass for GJ~614 to be $0.91 M_{\odot}$.


\subsection{GJ~649} \label{sec:gj649}

GJ~649 hosts a 0.33 $M_{Jup}$ planet in an eccentric 598.3-day orbit with a semi-major axis of 1.135 AU \citep{joh10}. We calculate GJ~649's circumstellar optimistic/conservative HZ's outer boundary to be 0.44 AU / 0.42 AU, putting the planet well beyond the outer edge of the system HZ, even at periastron. 

Our estimate for GJ~649's radius is $0.5387\pm0.0157 R_{\odot}$, which falls in the middle between the values in \citet[][$0.46^{+0.01}_{-0.02}  R_{\odot}$]{tak07}, \citet[][$0.49 R_{\odot}$]{zak79}, and \citet[][$0.616 \pm 0.026 R_{\odot}$]{hou10}.

Our effective temperature for GJ~649 is $3590\pm45$~K, which also falls in the middle of a fairly large range in literature values: \citet[][$3370$~K]{laf10b}, \citet[][$4185^{+161}_{-334}$~K]{amm06}, \citet[][$3717, 3782 \pm 58$~K]{sou10a}, \citet[][$3670$~K]{mor08}, and \citet[][$3503 \pm 50$~K]{hou10}.

Applying Equation~\ref{eq:massradius} to our radius value produces a mass value for GJ~649 of $0.54 M_{\odot}$. 
 

\subsection{GJ~876} \label{sec:gj876}

\begin{figure}										
\centering
\epsfig{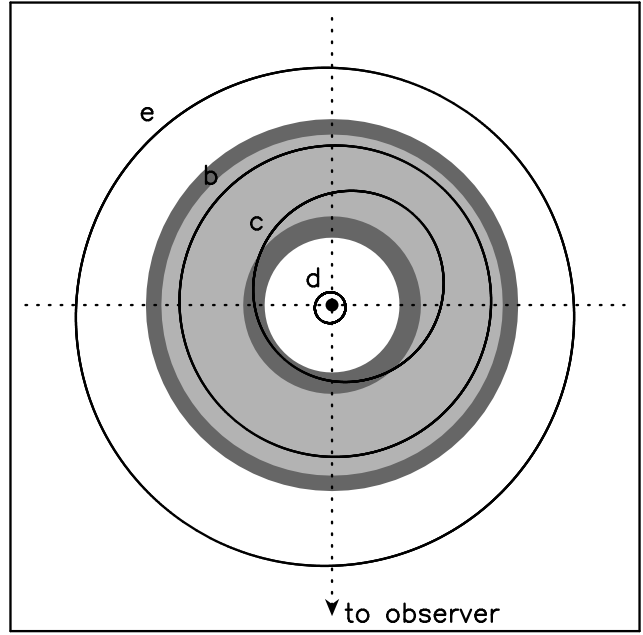} \\
\caption{Architecture of the GJ~876 system. The conservative HZ is shown in light grey, the optimistic HZ comprises both the light grey and dark grey regions. Planets b and c spend their entire orbits in the optimistic HZ. Planet b spends its entire orbit in the conservative HZ, whereas planet c spends 68.5\% of its orbital period in it. For details, see \S \ref{sec:properties}, \S \ref{sec:gj876}, and Table \ref{tab:parameters}. Orbital parameters for the planets are taken from \citet{riv10}. For scale: the size of the box is 0.8 AU $\times$ 0.8 AU.}
\label{fig:gj876_HZ}
\end{figure}

The late-type, multiplanet host GJ~876 has been studied extensively. There are four planets in orbit around the star \citep{cor10,riv10} at an inclination angle with respect to the plane of the system of $59.5^{\circ}$. The planet masses range from 6.83 $M_{\earth}$ to 2.28 $M_{Jup}$ in orbital distances that range from 0.02 to 0.33 AU with periods between 1.94 days and 124.26 days \citep{riv10}. Based on our optimistic / conservative HZ boundaries of 0.09--0.24 AU / 0.12--0.23 AU in Table \ref{tab:parameters}, two of the planets (b and c) are located within the system HZ. For an image of the system architecture, see Fig. \ref{fig:gj876_HZ}. 

We use the methods outlined in section 4 and equation 2 in \citet{von11c}, based on the work of \citet{skl07}, to calculate the equilibrium temperatures $T_{eq}$ for the GJ~876 planets via the equation

\begin{equation}\label{eq:equitemp}
  T_{eq}^4 = \frac{S (1 - A)}{f \sigma},
\end{equation}
where $S$ is the stellar energy flux received by the planet, $A$ is the Bond albedo, and $\sigma$ is the Stefan-Boltzmann constant \citep{skl07}

We differentiate between two scenarios, which are dependent on the efficiency of the heat distribution across the planet by means of winds, circulation patterns, streams, etc. The energy redistribution factor $f$ is set to 2 and 4 for low and high energy redistribution efficiency, respectively. Assuming a Bond albedo value of 0.3 the planetary equilibrium temperatures for $f$ = 4 are 587 K (planet d), 235 K (planet c), 186 K (planet b), and 147 K (planet e). The $T_{eq}$ values for $f$ = 2 are 698 K (d), 280 K (c), 221 K (b), and 174 K (e). These values scale as $(1-A)^\frac{1}{4}$ for other Bond albedo values (Equation \ref{eq:equitemp}). 

Previously estimated values for GJ~876's stellar radius are significantly below our directly determined value of $0.3761 \pm 0.0059 R_{\odot}$: $0.24 R_{\odot}$ \citep{zak79} and $0.3 R_{\odot}$ \citep{lau05,riv10}, the latter of which is the one that is frequently used in the exoplanet literature about GJ~876. In comparison to our value of $T_{\rm eff} = 3129\pm19$~K, literature temperatures for GJ~876 include a seemingly bimodal distribution of values: $3130$~K \citep{dod11}, $3165 \pm 50$~K \citep{hou12}, 3172~K \citep{jen09}, $3765^{+477}_{-650}$~K \citep{amm06}, and 3787~K \citep{but06}.

Finally, we derive a mass for GJ~876 of $M = 0.37 M_{\odot}$ using Equation~\ref{eq:massradius}.  


\subsection{HD~1461} \label{sec:hd1461}

HD~1461 hosts two super-Earths in close proximity to both the star and each other: a 7.6 Earth-mass planet in a 5.8-day orbit \citep{riv10a} and a 5.9 Earth-mass planet in a 13.5-day orbit \citep[][both are minimum masses]{may11}. Their semi-major axes are 0.063 and 0.112 AU, respectively, all well inside HD~1461's HZ, whose inner optimistic/conservative boundary is at 0.83 AU / 1.10 AU. 

Our radius estimate of $1.1987 \pm 0.0275 R_{\odot}$ is larger at the $> 2\sigma$ level than both radius estimates in the literature: $1.11 \pm 0.04 R_{\odot}$ \citep{tak07} and $1.0950 \pm 0.0260 R_{\odot}$ \citep{val05}. 

We measure an effective temperature for HD~1461 to be $5486\pm52$~K, which falls below all of the many temperature estimates in the literature for HD~1461 -- a sensible consequence given that our radius is larger than literature estimates.  The \citet{sou10a} catalog alone has 13 different $T_{\rm eff}$ values, ranging from 5683 to 5929~K. Additionally, we find temperature estimates of 5688~K \citep{hol09}, $5666 \pm 42$~K \citep{pru11}, and $5588 \pm 64$~K \citep{kol12} . 

We use the $Y^2$ isochrones following the method described in Section~\ref{sec:calculated} to derive and age and mass of HD~1461.  We obtain a mass of $0.94 M_{\odot}$ and an age of 13.8~Gyr. 


\subsection{HD~7924} \label{sec:hd7924}

The super-Earth ($M \sin i = 9.26 M_{\earth}$) orbits HD~7924 at a period of 5.4 days and at a semi-major axis of 0.057 AU \citep{how09}. The inner boundary of the optimistic/conservative HD~7924 system is at 0.47 AU / 0.62 AU from the star, well beyond the planetary orbit. 

The radius of HD~7924 has been estimated to be $R=0.78 \pm 0.02 R_{\odot}$ \citep{tak07}, which is identical to our direct measurement of $R = 0.7821 R_{\odot}$. The radius estimate of $R=0.754 R_{\odot}$ \citep{val05} is slightly below but consistent with our value. 

Our effective temperature measurement $T_{\rm eff} = 5075$~K falls into the middle of a large effective temperature range present in the literature for HD~7924: $4550$~K \citep{laf10b}, 4750~K \citep{wri03}, $5111^{+113}_{-128}$~K \citep{amm06}, 5121 to 5177~K (6 entries; \citealt{sou10a}), 5177~K \citep{pet11}, 5177~K \citep{val05}, $5153 \pm 5.8$~K \citep{kov04}, 5165~K \citep{mis08,mis12}.

Our derived mass from Equation~\ref{eq:massradius} is $M = 0.81 M_{\odot}$.


\subsection{HD~33564} \label{sec:hd33564}

\begin{figure}										
\centering
\epsfig{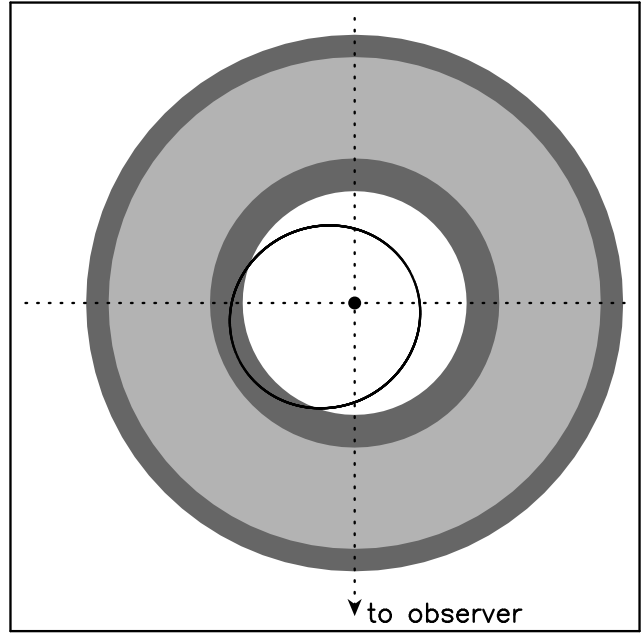} \\
\caption{Architecture of the HD~33564 system. The conservative HZ is shown in light grey, the optimistic HZ comprises both the light grey and dark grey regions. The planet in orbit around HD~33564 spends $\sim$ 43\% of its orbital period in the optimistic HZ. For details, see \S \ref{sec:properties}, \S \ref{sec:hd33564}, and Table \ref{tab:parameters}. Orbital parameters are from \citet{gal05}. The size of the box is 7 AU $\times$ 7 AU.}
\label{fig:hd33564_HZ}
\end{figure}

HD~33564 hosts a $M \sin i = 9.1 M_{Jup}$ planet in an eccentric 388-day orbit \citep{gal05} and an orbital semi-major axis of 1.1 AU. Since the orbital eccentricity is 0.34, its apastron distance is $r_{ap} = (1+e) a = 1.47$ AU. While the conservative HZ is located beyond HD~33564b's orbit, the planet spends around 43\% of its orbital duration inside the optimistic HZ, whose inner edge is at 1.31 AU. HD~33564's system architecture is shown in Figure \ref{fig:hd33564_HZ}.

We measure HD~33564's radius to be $1.4712\pm0.0219 R_{\odot}$.  This is consistent with the previous estimate based on SED fitting by \citet{van09} of $R = 1.45 \pm 0.03 R_{\odot}$. 

Our value for the effective temperature of HD~33564 is $6346\pm44$~K, which is largely consistent with the considerable number of estimates available in the literature: $T_{\rm eff} = 6440$~K \citep{wri03}, 6302~K \citep{mar95}, $6597^{+17}_{-708}$~K \citep{amm06}, $6307-6554$~K (3 entries; \citealt{sou10a}), $6250 \pm 150$~K \citep{but06}, $6531 \pm 70$~K \citep{van09}, 6456~K \citep{all99}, 6233~K \citep{sch09}, $6379 \pm 80$~K \citep{cas11}, 6307~K \citep{eir13}, 6394~K \citep{gra03}, 6250~K \citep{dod11}, and $6554 \pm 93$~K \citep{gon10}.

We estimate a mass of HD~33564 to be $1.31 M_{\odot}$ at an isochronal age of $2.2$~Gyr.  


\subsection{HD~107383 (= 11 Com)} \label{sec:hd107383}

The giant star HD~107383 has a substellar-mass companion in an eccentric 328-day orbit at a semi-major axis of 1.29 AU \citep{liu08}. Since HD~107383's luminosity is more than 100 times that of the sun, however, its optimistic/conservative HZ's inner boundary is at 8.19 AU / 10.83 AU, well beyond even the apastron of the known companion. 

Our radius estimate for the giant star HD~107383 is $15.78 \pm 0.34 R_{\odot}$ -- no previous radius estimates appear in the literature for this star. We measure an effective temperature for HD~107383 to be $4705\pm24$~K, which falls into the middle of previously published values:  4900~K \citep{wri03}, $4717^{+381}_{-283}$~K \citep{amm06}, 4690~K \citep{mcw90}, 4880~K \citep{hek07}, 4690~K, \citep{val04}, 4804~K \citep{sch07}, $4806 \pm 34$~K \citep{wu11}, 4690~K \citep{man09}, and 4873~K \citep{luc07}.

The evolutionary status and consequently the luminosity of this K0 giant are located outside of the range of the $Y^2$ isochrones. In addition, the star is evolved, and thus, Equation \ref{eq:massradius} is not applicable. We therefore cannot calculate its age or mass.





\subsection{HD~210702} \label{sec:hd210702}

HD~210702 hosts a $M \sin i = 1.9 M_{Jup}$ planet in a low-eccentricity, 355-day orbit \citep{joh07}. With a semi-major axis of 1.2 AU, the planet does not enter the system conservative or optimistic HZs, even at apastron.  

We measure HD~210702's radius and $T_{\rm eff}$ to be $5.2314\pm0.1171 R_{\odot}$ and $4780\pm18$~K, respectively, which is consistent with the interferometric (CHARA $K'$-band) values published in \citet[][$5.17\pm0.15 R_{\odot}$ and $4859\pm62$~K]{bai09}, as well as the radius estimated in the XO-Rad catalog of \citet[][$5.20\pm0.31 R_{\odot}$]{van09}. \citet{all99} quote 5.13 $R_{\odot}$ and 4897~K (with error estimates for all stars in their catalog of 6\% in radius and 2\% in $T_{\rm eff}$) -- also consistent with our direct values. Other estimates from \citet[][$4.45\pm0.07 R_{\odot}$ and $5510\pm44$~K]{joh07} and \citet[][$4.7 R_{\odot}$ and 4993~K]{mal13} are lower in radius and higher in effective temperature than our directly determined values.

Application of the $Y^2$ isochrones with input values from Table \ref{tab:parameters} for HD~210702 returns a stellar age of 5 Gyr and a mass of 1.29 $M_{\odot}$.


\section{Summary and Conclusion}    \label{sec:conclusion}

A very large fraction of the information on extrasolar planets that has been gathered over the course of the last 20 years is purely due to the study of the effects that the planets have on their respective parent stars. That is, the star's light is used to characterize the planetary system. In addition, the parent star dominates any exoplanet system as the principal energy source and mass repository. Finally, physical parameters of planets are almost always direct functions of their stellar counterparts. These aspects assign a substantial importance to studying the stars themselves: one at best only understands the planet as well as one understands the parent star. 

\begin{figure*}										%
\centering
\epsfig{file=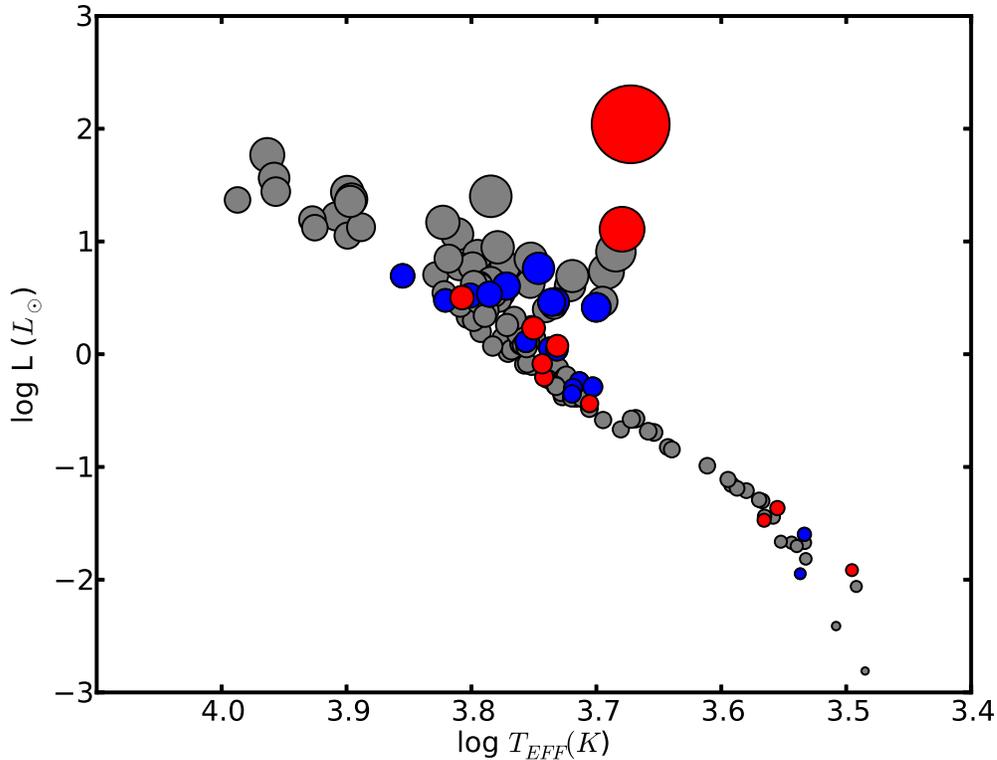,angle=0, width=15cm} \\
\caption{Empirical H-R Diagram for all main-sequence stars with interferometrically determined radii whose random uncertainties are smaller than 5\%, as published and compiled in \citet{boy12b,boy13}. The diameter of each data point is representative of the respective stellar radius. Error bars in effective temperature and luminosity are smaller than the size of the data points. Previously published exoplanet host stars are identified in blue \citep{ker03,bai08,bai09,van09,von11c,von11a,von12,boy13,hen13}. The exoplanet host stars that are presented in this work are shown in red. Stars that do not host any published exoplanets are shown in grey.}
\label{fig:ehs_hrd}
\end{figure*}

In this paper, we characterize eleven exoplanet host star systems with a wide range in radius and effective temperature, based on a 3.5-year long observing survey with CHARA's Classic beam combiner. For the systems with previously published direct diameters ($\rho$~CrB, GJ~614, and HD~210702), we provide updates based on increased data quantity and improved performance by the array. For the rest of the systems, only indirectly determined values for radius and effective temperature are present in the literature (if any exist at all). Our thus determined stellar astrophysical parameters make it possible to place our sample of exoplanet host stars onto an empirical Hertzsprung-Russell (H-R) Diagram. In Figure \ref{fig:ehs_hrd}, we show our targets along with interferometrically determined parameters of previously published exoplanet hosts and other main-sequence stars with diameter uncertainties of less than 5\%.

Due to the relatively low number of stars per spectral type in our sample, and due to the large variance among radius and temperature values quoted in the literature, it is impossible to quantify trends in terms of how indirectly determined values compare with direct counterparts as a function of spectral type, such as the ones documented in \citet{boy13}.  For the latest spectral type in our sample, and arguably the most interesting system in terms of exoplanet science, GJ~876, our directly determined stellar radius is significantly ($> 20\%$) larger than commonly used literature equivalents (\S \ref{sec:gj876}). 

We use our directly determined stellar properties to calculate stellar mass and age wherever possible, though the associated uncertainties are large for the KM dwarfs (\S \ref{sec:calculated}). Calculations of system HZ locations and boundaries, based on stellar luminosity and effective temperature, show that (1) GJ~876 hosts two planets who spend all or large parts of their orbital duration in the system HZ, whereas (2) the planet orbiting HD~33564 spends a small part of its period in the stellar HZ as its elliptical orbit causes it to periodically dip into it around apastron passage. 

CHARA's continously improving performance in both sensitivity and spatial resolution increasingly enables the direct measurements of stellar radii and effective temperatures further and further into the low-mass regime to provide comparison to stellar parameters derived by indirect methods, and indeed calibration of these methods themselves. 





\section*{Acknowledgments}

We thank the reviewer for the thorough analysis of the manuscript and helpful comments. We would furthermore like to express our sincere gratitude to Judit Sturmann for her tireless and invaluable support of observing operations at CHARA. Thanks to Sean Raymond, Barbara Rojas-Ayala, Phil Muirhead, Andrew Mann, Eric Gaidos, and Lisa Kaltenegger for multiple insightful and useful discussions on various aspects of this work. TSB acknowledges support provided through NASA grant ADAP12-0172. The CHARA Array is funded by the National Science Foundation through NSF grants AST-0606958 and AST-0908253 and by Georgia State University through the College of Arts and Sciences, as well as the W. M. Keck Foundation. This research made use of the SIMBAD and VIZIER Astronomical Databases, operated at CDS, Strasbourg, France (http://cdsweb.u-strasbg.fr/), and of NASA's Astrophysics Data System, of the Jean-Marie Mariotti Center \texttt{SearchCal} service (http://www.jmmc.fr/searchcal), co-developed by FIZEAU and LAOG/IPAG. This publication makes use of data products from the Two Micron All Sky Survey, which is a joint project of the University of Massachusetts and the Infrared Processing and Analysis Center/California Institute of Technology, funded by the National Aeronautics and Space Administration and the National Science Foundation. This research made use of the NASA Exoplanet Archive \citep{ake13}, which is operated by the California Institute of Technology, under contract with the National Aeronautics and Space Administration under the Exoplanet Exploration Program. This work has made use of the Habitable Zone Gallery at hzgallery.org \citep{kan12}. This research has made use of the Exoplanet Orbit Database and the Exoplanet Data Explorer at exoplanets.org \citep{wri11b}. This research has made use of the Exoplanet Encyclopedia at exoplanet.eu \citep{sch11}.




\bibliographystyle{mn2e}            

\bibliography{mn-jour,paper}      

\end{document}